\documentclass[11pt,a4paper]{emulateapj}
\usepackage{epsfig}
\usepackage{amsmath}
\usepackage{natbib, aas_macros}
\def \Simleq {\hspace{0.3em}\raisebox{0.4ex}{$<$}\hspace{-0.75em}\raisebox{-.7ex}{$\sim$}\hspace{0.3em}}
\def \Simreq {\hspace{0.3em}\raisebox{0.4ex}{$>$}\hspace{-0.75em}\raisebox{-.7ex}{$\sim$}\hspace{0.3em}}

\begin{document}

\title{The Formation of the massive galaxies in the SSA22 z=3.1 protocluster}
\author{M. KUBO\altaffilmark{1}, Y. K. UCHIMOTO\altaffilmark{1,2,3}, T. YAMADA\altaffilmark{1}, M. KAJISAWA\altaffilmark{4}, T. ICHIKAWA\altaffilmark{1}, Y. MATSUDA\altaffilmark{5}, M. AKIYAMA\altaffilmark{1}, T. HAYASHINO\altaffilmark{6}, M. KONISHI\altaffilmark{2}, T. NISHIMURA\altaffilmark{3}, K. OMATA\altaffilmark{3}, R. SUZUKI\altaffilmark{3}, I. TANAKA\altaffilmark{3},  T. YOSHIKAWA\altaffilmark{7}, D. M. ALEXANDER\altaffilmark{8}, G.G. FAZIO\altaffilmark{9}, J.-S. HUANG\altaffilmark{9}, B.D. LEHMER\altaffilmark{10} }
%
%
\altaffiltext{1}{Astronomical Institute, Tohoku University, 6-3 Aoba, Aramaki, Aoba-ku, Sendai, Miyagi, 980-8578, Japan}
\altaffiltext{2}{Institute of Astronomy, University of Tokyo, 2-21-1 Osawa, Mitaka, Tokyo, 181-0015, Japan}
\altaffiltext{3}{ Subaru Telescope, National Astronomical Observatory of Japan, 650 North A’ohoku Place, Hilo, HI 96720, USA }
\altaffiltext{4}{Research Center for Space and Cosmic Evolution, Ehime University, Bunkyo-cho 2-5, Matsuyama 790-8577, Japan}
\altaffiltext{5}{Chile Observatory, National Astronomical Observatory of Japan, Tokyo 181-8588, Japan}
\altaffiltext{6}{Research Center for Neutrino Science, Graduate School of Science, Tohoku University, Aramaki, Aoba-ku, Sendai, Miyagi, 980-8578, Japan}
\altaffiltext{7}{Koyama Astronomical Observatory, Kyoto Sangyo University, Motoyama, Kamigamo, Kita-ku, Kyoto 603-8555, Japan}
\altaffiltext{8}{Department of Physics, Durham University, Durham DH1 3LE, UK}
\altaffiltext{9}{Center for Astrophysics, Cambridge, MA 02138, USA}
\altaffiltext{10}{The Johns Hopkins University, Homewood Campus, Baltimore, MD 21218, USA}

\begin{abstract} The properties of $K$-band selected galaxies ($K_{\rm AB}<24$)
in the $z = 3.09$ SSA22 protocluster field are studied. $430$ galaxies at $2.6
< z_{\rm phot} < 3.6$ are selected as potential protocluster members in a 112
arcmin$^2$ area based on their photometric redshifts.  We find that $\approx
20$\% of the massive galaxies with stellar masses $>10^{11}$ $M_{\odot}$ at
$z_{\rm phot} \sim 3.1$ have colors consistent with those of quiescent galaxies
with ages $>$ 0.5 Gyr.  This fraction increases to $\approx 50\%$ after
correcting for unrelated foreground/background objects.  We also find that 30\%
of the massive galaxies are heavily reddened dusty star-forming galaxies.  Few
such quiescent galaxies at similar redshifts are seen in typical survey fields.
An excess surface density of 24$\mu$m sources at $z_{\rm phot} \sim 3.1$ is
also observed, implying the presence of dusty star-formation activity in the
protocluster. Cross-correlation with the X-ray data indicates that the
fraction of $K$-band selected protocluster galaxies hosting active galactic
nuclei (AGN) is also high compared with the field.  The sky distribution of the
quiescent galaxies, the 24$\mu {\rm m}$ sources, and the X-ray AGNs show
clustering around a density peak of $z=3.1$ Ly$\alpha$ emitters (LAEs).  A
significant fraction of the massive galaxies have already become quiescent,
while the dusty star-formation is still active in the SSA22 protocluster.
These findings indicate that we are witnessing the formation epoch of massive
early-type galaxies at the center of predecessors to present-day rich galaxy
clusters.

\end{abstract}



\section{Introduction}

The formation process of massive early-type galaxies is still an important and
open question to be understood. A tight color-magnitude relation has been
observed for galaxies in X-ray luminous clusters at $z \sim 1$ (e.g.,
\citealt{1997ApJ...483..582E}; \citealt{1998A&A...334...99K};
\citealt{2003ApJ...596L.143B}), implying that their major star formation
epochs took place at $z>2$. Since massive galaxies are the dominant populations
in the central regions of rich clusters in the local universe, overdense
regions in high redshift protoclusters provide important targets for directly
observing and studying the early phases of their formation.

\indent Several protoclusters at $z>2$ have now been identified (e.g.,
\citealt{1998ApJ...492..428S}; \citealt{2005A&A...431..793V};
\citealt{2005ApJ...620L...1O}; \citealt{2009MNRAS.395..114H};
\citealt{2009MNRAS.400L..66M};  \citeyear{2010MNRAS.403L..54M};
\citealt{2011MNRAS.416.2041M}; \citealt{2012ApJ...750..137T}) and density
excesses of red massive galaxies have been reported in some systems (e.g.,
\citealt{2006MNRAS.371..577K}; \citealt{2007MNRAS.377.1717K};
\citealt{2008ApJ...682..896K}; \citealt{2012ApJ...750..116U}). Among them, the
SSA22 protocluster at $z=3.09$ is one of the most outstanding structures. The SSA22 protocluster
was first discovered as a redshift spike of Lyman break galaxies (LBGs)
(Steidel et al. 1998) and then confirmed through the identification of a concentration of Ly$\alpha$ Emitters (LAEs; \citealt{2000ApJ...532..170S}; \citealt{2004AJ....128.2073H};
\citealt{2012AJ....143...79Y}). The density excess of LAEs in SSA22 is $\sim$15 times higher 
than that expected from mass fluctuations at $\sim 50$ Mpc scales
(\citealt{2012AJ....143...79Y}) and the rareness probability (i.e., the
probability to find such a rare overdensity) at 10 Mpc scale was found to be
0.0017\% (\citealt{2012ApJ...759..133M}). Thus, the SSA22 protocluster is well
characterized as a very significant density peak of galaxies at $z=3.09$.
Furthermore, overdensities of Ly$\alpha$ Blobs (LABs,
\citealt{2004AJ....128..569M}), ASTE/AzTEC 1.1mm sources
(\citealt{2009Natur.459...61T}; \citeyear{2010ApJ...724.1270T}) and active
galactic nuclei (AGN, \citealt{2009ApJ...691..687L};
\citeyear{2009MNRAS.400..299L}) have also been reported. 

\indent In order to study how stellar mass builds up in protocluster galaxies,
observations that probe rest-frame optical wavelength are essential. Uchimoto
et al. (\citeyear{2012ApJ...750..116U}, here after U12) conducted deep and
wide-field near-infrared (NIR) observations of the SSA22 protocluster using the
Subaru 8.2~m telescope equipped with the Multi-Objects InfraRed Camera and
Spectrograph (MOIRCS).  They observed galaxies to a magnitude limit of $K_{\rm
AB}=24$ over a $111.8$~arcmin$^2$ area around the LAE density peak.  They
selected protocluster galaxy candidates using photometric redshifts and
simple color criteria appropriate for Distant Red Galaxies (DRGs,
\citealt{2003ApJ...587L..79F}) and for Hyper Extremely Red Objects (HEROs,
\citealt{2001ApJ...558L..87T}).  Through these selection criteria, they found
that there is a significant surface number density excess of these 
stellar mass selected galaxies compared to those in blank fields. 

\indent Another notable discovery by U12 was the detection of ``multiple
merging galaxies" associated with the LABs and an ASTE/AzTEC sub-mm source.
They found that multiple sub-components with $z_{\rm phot}$ $\approx
3.1$ were clustered with separations up to 20$''$ (150 kpc at $z=3.09$) within
extended Ly$\alpha$ halos of LABs or within the beam size of the ASTE/AzTEC.
This suggested that star formation is actively occurring in each sub-component
before these systems undergo mergers that subsequently lead to the
formation of massive galaxies.  High-resolution cosmological numerical
simulations show that such phenomena are indeed expected in the early phase of
the formation of massive early-type galaxies (e.g.,
\citealt{2003ApJ...590..619M}; \citealt{2007ApJ...658..710N}).

\indent In this paper, we investigate the properties of galaxies in the SSA22
protocluster in more detail not only by using the same MOIRCS NIR data as used
in U12 but also by using the {\it Spitzer} IRAC 3.6--8.0$\mu$m and MIPS 24$\mu$m
data, as well as the {\it Chandra} 0.5--8 keV X-ray data.  With the IRAC data,
we can constrain the spectral energy distributions (SEDs) of galaxies over
a wider wavelength range, enabling us not only to obtain more accurate
photometric redshifts but also to conduct improved stellar population analyses
of the protocluster galaxies.  We make use of the 24$\mu$m data to constrain
the dust emission and the star-formation and/or AGN activities.

\indent Association of AGNs with the galaxies in the protocluster is also an
important issue.  In the SSA22 region, \citeauthor{2009ApJ...691..687L}
(\citeyear {2009ApJ...691..687L}; \citeyear{2009MNRAS.400..299L}) investigated
the AGN hosting rate among LAEs and LBGs and found an excess AGN fraction in
the protocluster over that of the field.  Since there is a tight correlation
between black hole mass and the spheroid stellar mass of the host galaxies in
local universe (e.g., \citealt{1998AJ....115.2285M}), we can investigate at
what level the excess AGN fraction found in protocluster galaxies is due to an
enhancement in the black hole masses of these galaxies.

\indent The paper is organized as follows.  Observations and data reduction are
described in Section 2. Our techniques for selecting protocluster galaxies based on
photometric redshifts are described in Section 3. In Section 4, we investigate the
properties of the protocluster galaxy candidates.  We discuss the detailed
stellar populations of the galaxies in Section 5.  In this paper, we assume
cosmological parameters of $H_0 = 70$ km s$^{-1}$ Mpc$^{-1}$ , $\Omega_{M} =
0.3$ and $\Omega_{\Lambda} = 0.7$.  $E(B-V)=0.08$ is used for the Galactic
extinction in the SSA22 field following \citet{2012AJ....143...79Y}.  We use
the AB magnitude system throughout this paper. 

\section{Observations \& Data reduction}

\subsection{Description of data} \indent The $JHK_s$-band images were obtained
by using MOIRCS equipped on Subaru 8.2 m telescope on June and August 2005,
July 2006, and September 2007 (U12). The area observed with MOIRCS is located
in the SSA22-Sb1 field (\citealt{2004AJ....128..569M}), which covers the
highest density region of the LAEs at $z\approx3.09$
(\citealt{2004AJ....128.2073H}; \citealt{2012AJ....143...79Y}). The whole
observed area consists of three 4$'$ $\times$ 7$'$ fields of views (FoVs; M1,
M2, and M6) and three 4$'$ $\times$ 3.5$'$ FoVs (M3, M4, and M5; See also Fig.
1 in U12). In total, 111.8 arcmin$^2$ area of the sky was observed.  The data
was calibrated to the UKIRT $JHK$-band photometric system
(\citealt{2002PASP..114..180T}). A detailed description of the data and the
reduction procedures are described in \citet{2008PASJ...60..683U} and U12.
Briefly, the limiting magnitude in the $K$-band is $K = 24.1$ (5$\sigma$
detection limit for a 1.1$''$ diameter aperture).  The full width at half
maximum (FWHM) of the point spread function (PSF) in the $K$-band is $\sim$
0.5$''$. The sizes of PSF in $J$-band are similar to those in $K$-band in the
M1, M2, M3 and M6 fields, while those in M4 and M5 fields are 0.6$''$ and
0.7$''$, respectively. Aperture photometry was performed using the PHOT task in
IRAF.

\indent Source detections were performed on the $K$-band images using
SExtractor (\citealt{1996A&AS..117..393B}).  Objects with 16 continuous pixels
above 1.5$\sigma$ of the background fluctuation were extracted.  We use
MAG\_AUTO of SExtractor as the pseudo total magnitudes of the objects.
Hereafter, this paper is focused on the sample of the $K$-band selected
galaxies with $K_{\rm AUTO}<24$.

\indent To supplement the $K$-band data, we use the $u^{\star}$-band archival
image taken by CFHT MegaCam (P.I.  Cowie, see also
\citealt{2004AJ....128..569M}), the $BVRi'z'$ and the $NB497$-band images taken
by Subaru Suprime-Cam (\citealt{2004AJ....128..569M};
\citealt{2004AJ....128.2073H}) and the 3.6$\mu$m, 4.5$\mu$m, 5.8$\mu$m and
8.0$\mu$m-band images by {\it Spitzer} IRAC (\citealt{2009ApJ...692.1561W}).
The entire field observed with MOIRCS is covered by the optical and the IRAC
images. $NB497$ is a narrow-band filter with the central wavelength of 4977\AA,
which is used to detect Ly$\alpha$ emission from galaxies at $z=3.061-3.125$.
All the optical images are smoothed so that the PSF sizes are matched to be
1.0$''$ in FWHM. The PSF sizes of the IRAC images are $\sim 1.7''$, slightly
depending on the wavelength. Hereafter, IRAC 3.6$\mu$m-band AB magnitudes are
quoted using the [3.6$\mu$m] notation; we apply the same corresponding
notations to the 4.5, 5.8, and 8.0$\mu$m bands.

\indent  We also made use of the {\it Spitzer} MIPS 24$\mu$m data
(\citealt{2009ApJ...692.1561W}), the point source catalog of {\it Chandra}
0.5-8 keV data (\citealt{2009ApJ...691..687L}), the archived {\it Hubble Space
Telescope} ({\it HST}) {\it Wide Field Camera 3 } (WFC3) $F110W$- and
$F160W$-band data, and the {\it Advanced Camera for Survey} (ACS) $F814W$-band
data (proposal ID; 11636, PI: Siana, Exposure time for each image are 2611
$s$ for $F110W$- and $F160W$-band, respectively, and 6144 $s$ for
$F814W$-band). 

\indent The MOIRCS field is almost entirely covered by the 24$\mu$m data except
for $\sim4$ arcmin$^2$ region at the north-west edge of the M5 field.  The {\it
Chandra} data covers the entire MOIRCS field except for the M5 region.  We
therefore study the X-ray properties of galaxies in the available 99.8
arcmin$^{2}$ overlapping area.  The nominal limiting fluxes are $\approx
6\times10^{-16}$ erg cm$^{-2}$ s$^{-1}$ in the full-band (0.5-8 keV),
$\approx2\times10^{-16}$ erg cm$^{-2}$ s$^{-1}$ in the soft-band (0.5-2 keV)
and $\approx1\times10^{-15}$ erg cm$^{-2}$ s$^{-1}$ in the hard-band (2-8 keV),
respectively.  The {\it HST} images cover only a few patchy regions, less than
10 arcmin$^2$ total area (for $F160W$-band), but are useful to see the
structural properties of some interesting objects.  The data used in this
article are summarized in Table~\ref{tab:tab1}.\\

\subsection{Optical and NIR photometry} 

\indent We use photometric redshifts obtained from SED fitting to select
protocluster galaxy candidates.  For this purpose, we use the flux of the
objects in $u^{\star}BVRi'z'JHK$ and IRAC 3.6$\mu$m, 4.5$\mu$m, 5.8$\mu$m and
8.0$\mu$m bands after correcting for the effects of different PSF sizes. The
flux in the optical $u^{\star}BVRi'z'$ bands is measured using a
2.0$''$-diameter circular aperture.  We smoothed the $JHK$ band images to have
the PSF sizes of 1.0$''$ and then obtain 2.0$''$-diameter aperture fluxes.
Since the IRAC images have large PSF sizes ($\sim 1.7''$ diameter), we first
obtain 3.0$''$-diameter fluxes and then apply aperture corrections.  To obtain
the aperture correction factors for each object, we smoothed the $K$-band
images to have the PSF sizes of 1.0$''$ and 1.7$''$ to be matched with the
optical and the IRAC images, respectively.  For each object, the ratio of the
fluxes within a 2.0$''$-diameter aperture on the former image to the flux
within a 3.0$''$-diameter aperture on the latter image was measured to obtain
the correction factor.

\indent We also use $J-K$, $i'-K$ and $K-$[4.5$\mu$m] colors of the galaxies in
our analyses.  The $J-K$ colors are measured from 1.1$''$-diameter apertures
after the $J$ and $K$ images are smoothed so that their PSF sizes are matched
with each other. If objects are not detected in $J$-band above 2$\sigma$
threshold, we use the 2$\sigma$ values for upper limits. The $i'-K$ colors are
the same as those obtained for SED fitting (see description of photometry
above).   To obtain $K-[4.5$$\mu$m] colors, we measured the 3.0$''$-diameter
aperture flux on the IRAC images and then corrected them to the
1.1$''$-diameter aperture values.  The aperture correction factor is the ratio
of the 1.1$''$-aperture flux on the original $K$-band images to the
3.0$''$-aperture flux on the smoothed $K$-band image.

\subsection{MIPS 24$\mu$m photometry} 

Since the PSF size of the
24$\mu$m image is relatively large ($\sim 6.0''$), source confusion is very
significant.  Therefore, we obtained 24$\mu$m fluxes using the PSF fitting method described in \citet{2005ApJ...632..169L}, which makes use of the
DAOPHOT package in IRAF (\citealt{1987PASP...99..191S}).
The procedure starts by fitting a template PSF, derived using bright
sources, to all the sources in the image.  
Next, source fluxes are estimated 
by subtracting the contributions from blended sources.  For this task, we used the positions
of objects detected in the $K$-band images as a prior.  Fluxes were obtained using 6.0$''$-diameter apertures, which were aperture corrected using the PSFs derived from isolated bright sources.

Due to the inhomogeneous exposure time over the observed field, the noise level
of the 24$\mu$m image varies by position.  The background noise level expected
for a 6.0$''$ diameter aperture was estimated within 80$'' \times$ 80$''$
square regions surrounding each source. The limiting total flux (3$\sigma$) is
$\approx$ 40--100 $\mu$Jy.  We use the conservative nominal detection limit of
$f_{24\mu{\rm m}}=100$ $\mu$Jy in investigating their sky distribution and
their surface number density. 

\begin{figure}[ht] 
\begin{center}
\includegraphics[width=130mm]{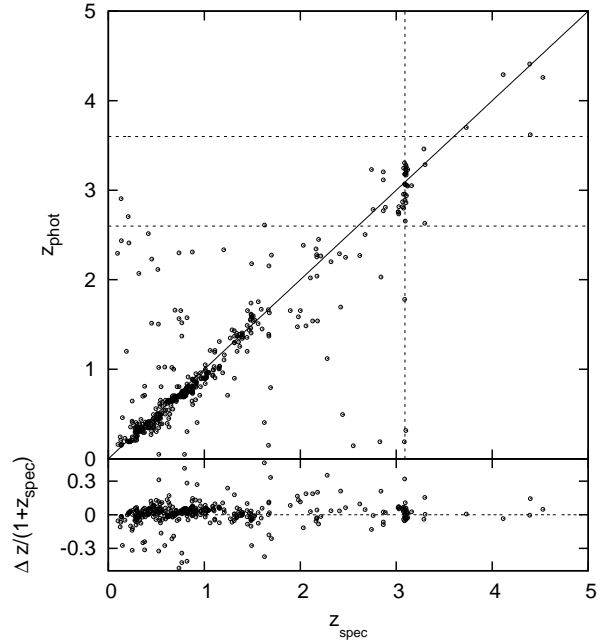}
\caption{ {\it Top}; The comparison between the spectroscopic redshifts
available in the NASA/IPAC Extra-galactic Database and our photometric
redshifts of the $K$-selected galaxies in the SSA22 field. The vertical dashed
line shows the median redshift of the protocluster. The horizontal dashed lines
are drawn at $z_{\rm phot}=2.6$ and $z_{\rm phot}=3.6$. We plot the relative
errors of the photometric redshifts $\Delta z/(1+z_{\rm spec})$ where $\Delta
z= (z_{\rm spec}-z_{\rm phot})$ in the {\it bottom} panel. 
} 
\label{fig:photz}
\end{center} 
\end{figure}
\begin{figure*} 
\begin{center} 
\leavevmode
\epsfxsize=16cm\epsfbox{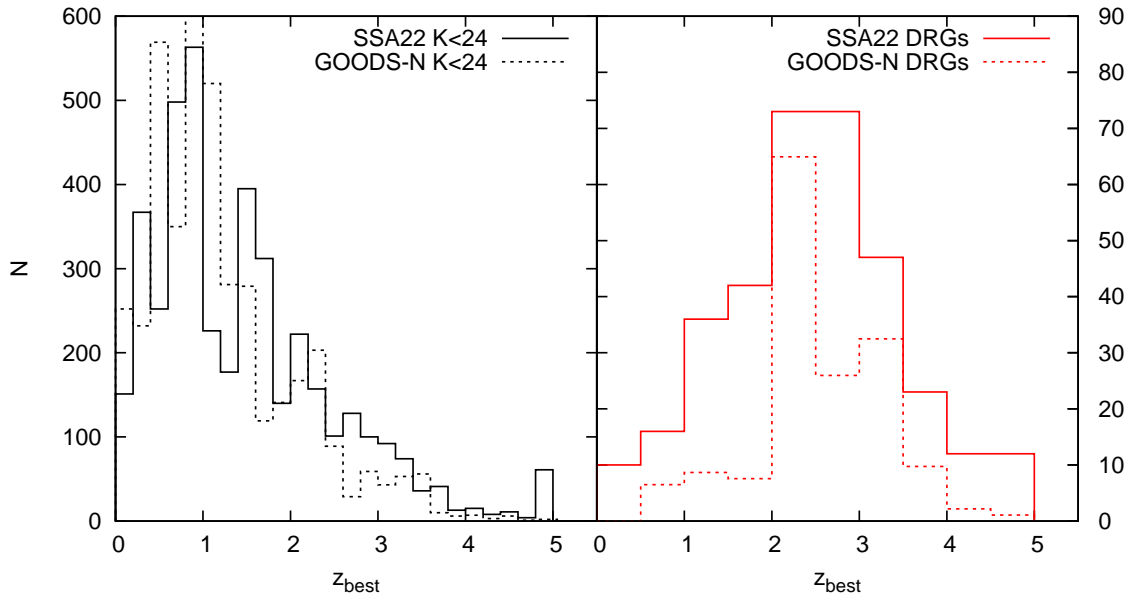}
\caption{{\it Left}; The $z_{\rm best}$ distributions of the $K$-selected
galaxies. All the galaxies with $K<24$ are plotted. The black solid line is the
redshift distribution of the $K$-selected galaxies in the SSA22 field and the
dashed line is that in GOODS-N field. {\it Right}; The $z_{\rm best}$
distributions of the DRGs with $K<24$. The red solid line is the redshift
distribution of the DRGs in the SSA22 field and the dashed line is that in
GOODS-N field. } 
\label{fig:redshift-dist} 
\label{fig:disk} 
\end{center}
\end{figure*}

\section{The photometric redshift selection}

\subsection{The photometric redshift}

In total, we detected $\approx$ 4500 objects with $K<24$ in the SSA22 MOIRCS
images (see description of source searching procedure in Section 2.1).  Using
this catalog of sources, we identified protocluster candidates based on the
photometric redshifts ($z_{\rm phot}$), which were measured using the following
procedure.

\indent We derived photometric redshifts for each object using the SED fitting
code HyperZ (\citealt{2000A&A...363..476B}).  We used the fluxes in the
$u^\star BVRi'z'JHK$, 3.6$\mu$m, 4.5$\mu$m, 5.8$\mu$m and 8.0$\mu$m bands as
input to HyperZ.  The fluxes used as input are the PSF-corrected values
described in the previous section.  For the SED templates, we used GALAXEV
(\citealt{2003MNRAS.344.1000B}) models with the Salpeter initial mass function
(IMF) and the range of the metallicity from 1/100 to 1 times the solar value.
The star formation histories of the templates are single burst, constant-rate
continuous star formation, and exponentially decaying star formation with
timescales of $\tau = 0.1-30$ Gyr.  We adopt the \citet{2000ApJ...533..682C}
extinction law with a range of extinctions covering $E(B-V)= 0.0-2.0$.  A total
of $\sim 200$ objects with $K<20$ that are flagged as stars by SExtractor are
removed from the sample.  When the galaxies are significantly blended with
adjacent objects only in the IRAC images, we obtain the photometric redshifts
after excluding the IRAC data. 

\indent In the HyperZ program, photometric redshifts are determined by the best
fit SED model based on $\chi^2$ minimization.  Objects with $\chi_{\nu}^2 >
100$ ($\chi_{\nu}^2 = \chi^2/\nu$; $\nu=12$) where $\nu$ is the number of
degrees of freedom, are not used in the following analysis due to the poor
goodness of fit.  The catastrophic failure fraction rapidly increases above
this threshold.  In total, we excluded 134 objects from our sample that fail
this criterion; therefore, 97\% of the original $K$-selected galaxy sample are
included in our further analyses.

\indent The top panel of Fig.~\ref{fig:photz} shows the comparison of our
photometric redshifts of $K$-selected galaxies with spectroscopic redshifts
($z_{\rm spec}$) available from the NASA/IPAC Extra-galactic Database (NED).
There are 456 objects with spectroscopically confirmed redshifts.  Out of these
456 objects, 33 have $z_{\rm spec}=3.02$--$3.16$ and are likely protocluster
sources.  We excluded five known QSOs and an additional three QSO-like point
sources from this plot.  The latter objects have not been classified as QSOs in
previous studies, but their SEDs are different from the galaxy templates and
more closely resemble those of QSOs.  The bottom panel of Fig.~\ref{fig:photz}
shows the relative errors of the photometric redshifts, $\Delta z/(1+z_{\rm
spec})$ where $\Delta z= (z_{\rm spec}-z_{\rm phot})$. The standard deviation
is $<\Delta z/(1+z_{\rm spec})> = 0.08$, excluding the galaxies with $|\Delta
z|/(1+z_{\rm spec}) > 0.5$. For galaxies at $z_{\rm spec}=3.02-3.16$, $<\Delta
z/(1+z_{\rm spec})>$ is 0.07.  The median photometric redshift of the
protocluster galaxies is $z_{\rm phot} = 3.06$, which agrees well with that of
the spectroscopic sample.  On the other hand, there are three catastrophic
failures for the $z_{\rm spec} = 3.02-3.16$ objects.  We closely inspected
their SEDs and found that these three failures were likely due to the
misidentifications of Lyman break and Balmer break features, a strong [O$_{\rm
III}$]$\lambda\lambda$5007 emission line, and a weak Lyman break.

\indent In U12, the photometric redshifts were obtained with $<\Delta
z/(1+z_{\rm spec})>= 0.12$ for the whole $K$-selected galaxies and $<\Delta
z/(1+z_{\rm spec})> = 0.08$ for these at $z_{\rm spec}\approx3.09$. We have
thus improved the photometric redshift measurement over those provided by U12.  

\indent In the following analyses, we define our best redshift measurement,
$z_{\rm best}$, as the spectroscopic redshift when its available; otherwise, we
adopt the photometric redshift as $z_{\rm best}$.  In total, we select 433
$K$-selected galaxies at $2.6<z_{\rm best}<3.6$, which may contain a
significant fraction of protocluster galaxies at $z\approx3.09$.  A breakdown
of the number of $K$-selected galaxies with spectroscopic and photometric
redshifts is shown in Table \ref{table:table-nc}.  Note that we removed 10
objects which are blended with nearby very bright objects in the optical
images.

\subsection{The redshift distribution of the $K$-selected galaxies}

\indent The left panel of Fig.~\ref{fig:redshift-dist} is the $z_{\rm best}$
distributions of the $K$-selected galaxies in the SSA22 field.  For comparison,
we also plot the redshift distribution of the $K$-selected GOODS-N galaxies from the
MOIRCS Deep Survey (MODS) catalog (\citealt{2006PASJ...58..951K};
\citeyear{2011PASJ...63S.379K}).  In this comparison, the number of galaxies in
GOODS-N has been normalized to the area of the SSA22 survey field.  MODS
obtained photometric redshifts of the galaxies with $K < 25.1$ $(5\sigma$
detection limit) in a 103.3 arcmin$^2$ area.  The rms error of the photometric
redshifts are $<\Delta z /(1+z_{\rm spec})> \sim 0.12$ for the whole sample and
$\sim 0.08$ for the sample at $z \sim 3$ after clipping the catastrophic
failures with $|\Delta z | /(1+z_{\rm spec})> 0.5$
(\citealt{2007PASJ...59.1081I}).  Note that the rms error of the photometric
redshifts in the SSA22 field is comparable to that of MODS but contains more
catastrophic failures.  That is $\approx$ 1\% of the objects at $z_{\rm
spec}<2$ in our SSA22 catalog are contaminants at $2.6<z_{\rm phot}<3.6$, while
the equivalent contaminant fraction for the MODS is $\sim$ 0.2\%.  The entire
redshift distributions agree well while there is a notable excess in the number
of $z\sim 3$ sources in the SSA22 field. 

\indent The right panel of Fig.~\ref{fig:redshift-dist} displays the $z_{\rm
best}$ distributions of DRGs, which are selected using the color criterion $J-K
>1.4$.  DRGs are thought to be dust obscured star-forming galaxies and/or
relatively old passive galaxies with significant Balmer or 4000\AA\quad breaks
at $2<z<4$ (\citealt{2003ApJ...587L..79F}).  Spectroscopic observations show
that DRGs with $K > 21$ are indeed dominated by galaxies at $z > 2$ (e.g.
\citealt{2007ApJ...655...51W}; \citealt{2011PASJ...63S.379K}).  DRGs in both
SSA22 and GOODS-N are dominated by galaxies at $z_{\rm best}>2$. In SSA22, we
detected 364 DRGs among the entire galaxies with $K<24$ and 118 of them are at
$2.6<z_{\rm best}<3.6$.  From Fig.~\ref{fig:redshift-dist} it is clear that
there is an excess number density of DRGs in SSA22 compared with GOODS-N over
the wide range of redshift.  The density excess at $2<z<4$ is likely to be
caused by the excess number of galaxies in the $z=3.09$ protocluster.  Indeed,
5 of the 9 DRGs in SSA22 with known spectroscopic redshifts are at $3.09<z_{\rm
spec}<3.12$.  The density excess at $z<2$ may be due to a true overdensity at
$1<z<2$; however, it is possible that high-redshift ($z > 2$) contaminants with
catastrophic photometric redshift failures may contribute to the low-redshift
overdensity.  The scatter of photometric redshift is generally larger for the
featureless reddened spectra.  

\begin{figure*} \begin{center} \leavevmode
\epsfxsize=14cm\epsfbox{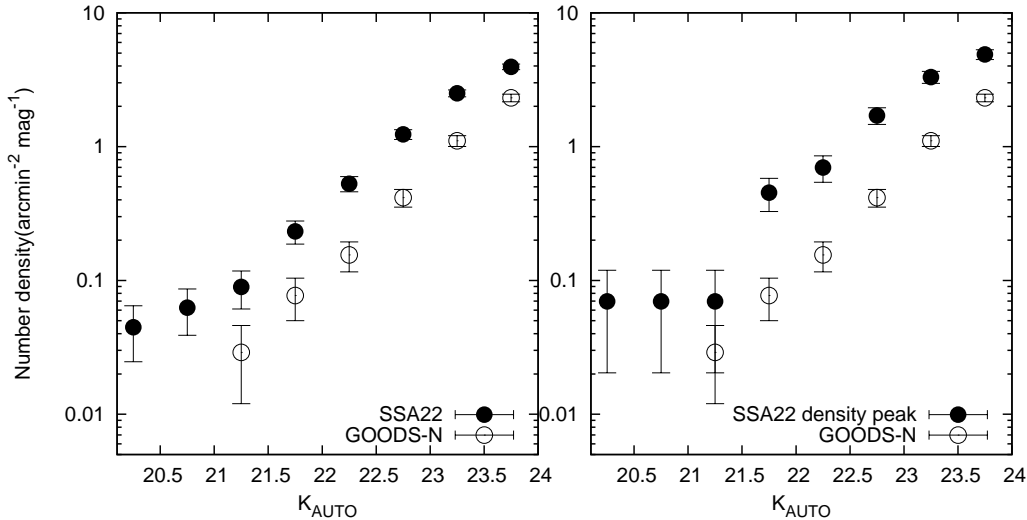}
\caption[Numberdensity]{{\it Left}; The cumulative surface number density of
the $K$-selected galaxies at $2.6<z_{\rm best}<3.6$. The black filled circles
are that in the entire SSA22 field and the black open circles are that in the
GOODS-North field. {\it Right}; Similar to the left panel but the black filled
circles are the cumulative surface number density at the SSA22 highest density
region. } \label{fig:count-k} \end{center} \end{figure*}

\subsection{The density excess of the $K$-selected galaxies at $2.6<z_{\rm best}<3.6$}

In this subsection, we discuss further the surface number density excess of the
$K$-selected galaxies at $2.6<z_{\rm best}<3.6$ in the SSA22 field. Note that
the excess of the purely color-selected DRGs (i.e., no photometric redshift
selection) was previously reported by U12. 

\indent In the left panel of Fig.~\ref{fig:count-k}, we show the cumulative
surface number density of the $K$-selected galaxies at $2.6<z_{\rm best}<3.6$
in the entire SSA22 field. The error bar in each bin corresponds to the
1$\sigma$ Poisson error. We also plot the distribution in GOODS-N for
comparison. The surface number density of the galaxies with $K < 24$ in the
SSA22 field is 3.87$\pm0.19$ arcmin$^{-2}$, which is 1.7 times of that in
GOODS-N. If we subtract the contribution of field galaxies (based on GOODS-N)
from the SSA22 number counts, we estimate that there is an excess of $\approx
180$ $K$-selected galaxies, potentially sources that are members of the
protocluster. 

\indent We identified the position of the large-scale density peak of the
$K$-selected galaxies at $2.6<z_{\rm best}<3.6$ on the sky by smoothing their
distribution with a Gaussian kernel of scale $\sigma=0.5$~arcmin.  We found the
location of the peak at ($\alpha$, $\delta$) $\approx$ (22${\rm h}$17$^{\rm
m}$20$^{\rm s}$, +00$^\circ$17.$'$7), which is indicated in Fig.
\ref{fig:dis-passive}.  Interestingly, the location of the peak is within
1~arcmin of the peak of the $z = 3.09$ LAE distribution, which was obtained
using a smoothing scale of 1.5 arcmin in \citet{2012AJ....143...79Y}.   The
density peak of the simple color-selected DRGs was also found to be near that
of the protocluster LAEs (U12). The right panel of Fig.~\ref{fig:count-k} is
the cumulative surface number density of $K$-selected galaxies at $2.6<z_{\rm
best}<3.6$ within 1.5 Mpc radius around the density peak.  We find the density
of $K < 24$ galaxies in this region of SSA22 to be 2.1 times that found in the
GOODS-N field. 

\indent In order to test whether the $K$-selected source overdensity in SSA22
over GOODS-N was due to differences in accuracy of photometric redshifts, we
performed simulations.  While the rms errors after clipping the outliers are
comparable, there are more catastrophic failures ($|\Delta z| \Simreq 0.5$) in
the SSA22 sample.  We evaluated the relative fraction of the catastrophic
failures in the SSA22 sample compared to the MODS sample by constructing mock
catalogs where the distributions of the parameters in SED fitting are similar
to those in the MODS sample.  We added photometric errors to the mock catalog
sources and obtained the fraction of the catastrophic failures.  In this
process the mock photometric error distribution was matched to those obtained
for the SSA22 galaxies.
After 100 realizations, we found that 1.0$\pm$0.2\% of the galaxies at
$z_{\rm model} < 2$ are contaminated by the $2.6<z_{\rm phot}<3.6$ sample in
the mock catalogs.  This contamination fraction is only 0.2\% for the real MODS catalog. This
implies that at most $\sim$ 1.1 times of the surface number density
excess can be attributed to contamination due to the catastrophic redshift measurements.
Thus, the density excess of the $K$-selected galaxies at $2.6 <z_{\rm
phot}<3.6$ in the SSA22 field is not caused by the artifact due to the
difference in accuracy of the photometric redshift. 

\indent U12 reported that the surface number density of the pure color-selected
DRGs (i.e., without photometric redshift constraint) is 1.8--2.1 times that in
the general field.  To extend their analysis, we computed the number density of
DRGs with $2.6<z_{\rm best}<3.6$.  We corrected the detection completeness of
the DRGs in the M4 and M5 field where the images are slightly shallow due to
the relatively large PSF sizes.  The correction factors at the faintest bins
are found to be $\approx 1.1$.  After applying these corrections, we obtained
the surface number density of the DRGs at $2.6<z_{\rm best}<3.6$.  At $K<24$,
we find a surface density of $1.04\pm0.10$ arcmin$^{-2}$, which is 2.5 times of
that found in GOODS-N.  This excess increases to a factor of 2.7 near the
$K$-selected source density peak. 

\indent Overdensities of red galaxies in protoclusters at $z\sim2$ have also
been found in other studies (e.g., \citealt{2007MNRAS.377.1717K};
\citealt{2007ApJ...656...66Z}; \citealt{2010A&A...509A..83D}). On the other
hand, no significant density excesses of the DRG-like objects were observed in
some protoclusters at $z\sim3$, including MRC 0943-242 at $z=2.93$
(\citealt{2010A&A...509A..83D}) and MRCS0316-257 at $z=3.13$
(\citealt{2010MNRAS.405..969K}; based on Balmer-break selection of galaxies
with colors similar DRGs). \citet{2010MNRAS.405..969K} noted that it is
difficult to detect the surface number density excess of DRGs with their sample
due to limited area and depth.  We note that our NIR data in the SSA22 field is
wider and deeper than those used by \citet{2010A&A...509A..83D} and
\citet{2010MNRAS.405..969K}.

\indent A density excess of 24$\mu$m detected galaxies at $2.6<z_{\rm
best}<3.6$ was also found to be 1.9 times of that of GOODS-N.  As we will
discuss below, the 24$\mu$m detected galaxies at $z\sim3$ are likely to be
heavily dust obscured star-forming galaxies and/or AGNs.  At such high
redshifts, the 24$\mu$m detection limit corresponds to an SFR of several
100--1000 $M_{\odot}$yr$^{-1}$.  The density excess of 24$\mu$m sources around
high redshift radio sources (HzRGs) has been reported in other fields.  For
example, \citet{2012A&A...539A..33M} reported an average overdensity of
2.2$\pm$1.2 of 24$\mu$m sources ($<0.3$ mJy) using simple source counts within
1.75$'$-radius regions around HzRGs at $1\leq z \leq 5.2$. Also,
\citet{2011PASJ...63S.415T} reported a density excess around a HzRG at
$z=2.48$.  

\section{Properties of $K$-selected galaxies at $2.6<z_{\rm best}<3.6$} In this
section, we investigate the properties of the $K$-selected galaxies with
redshifts $2.6<z_{\rm best}<3.6$.  For our purposes here, we compute stellar
masses and SFRs for these galaxies assuming they are members of the
protocluster and located at $z=3.09$.  With this assumption, we repeated the
SED fitting analysis described in the previous section with the fixed redshift,
$z=3.09$. 

\begin{figure*}[ht] 
\begin{center} 
\leavevmode
\epsfxsize=17cm\epsfbox{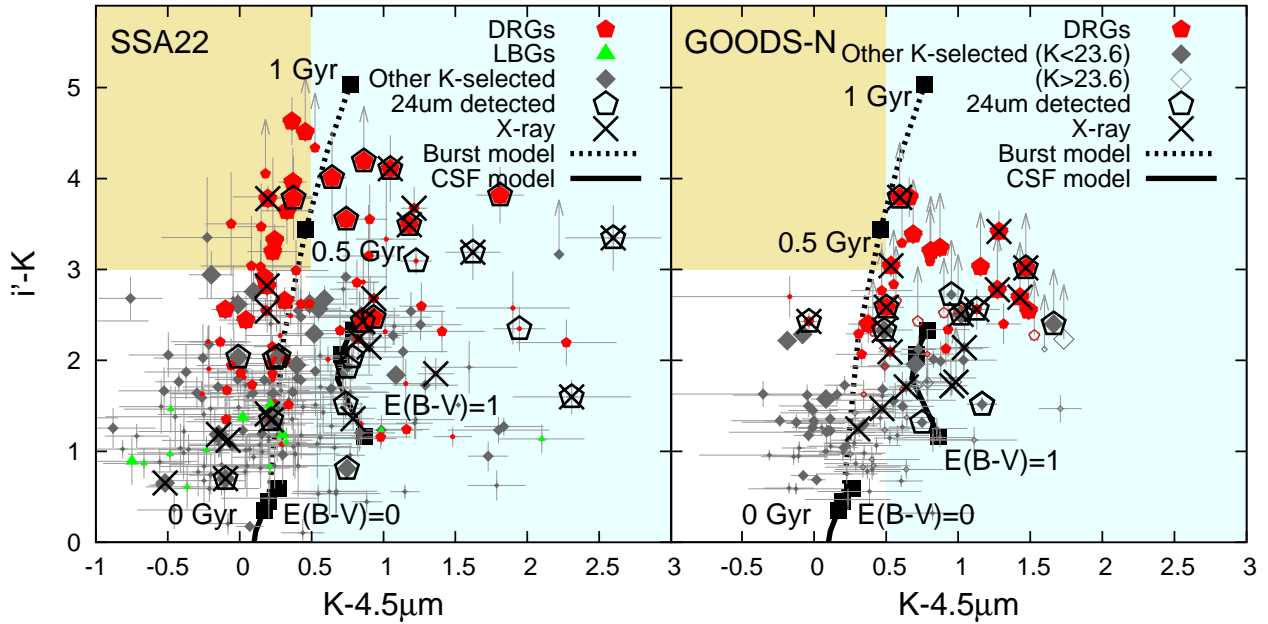}

\caption[i-K v.s. K-4.5um]{ $i'-K$ versus $K-[4.5\mu$m] two color diagram of
the $K$-selected galaxies at $2.6<z_{\rm best}<3.6$ in the SSA22 (left) and
GOODS-N (right) field. The gray filled diamonds are all the $K$-selected
galaxies which are also detected in [$4.5\mu$m]. The size of the symbols
reflects the stellar masses of the objects classified as $M_{\rm
star}<10^{10.5}$ $M_{\odot}$ (small), $10^{10.5}$ $M_{\odot} <{\rm M}_{\rm
star} <10^{11}$ $M_{\odot}$ (medium) and $M_{\rm star}>10^{11}$ $M_{\odot}$
(large). The red filled pentagons and the green filled triangles are the
$K$-selected galaxies at $2.6<z_{\rm best}<3.6$ classified as DRGs and LBGs,
respectively. The open black pentagons are the galaxies detected in 24$\mu$m
and the black crosses are those detected in {\it Chandra} full- and/or soft-
and/or hard-bands. The black dotted line and square points are the color
evolution track with age for single burst star formation model at $z=3.1$. The
square points are at 0, 0.5, 1.0 Gyr from bottom to top. The black solid lines
and square points are those for constant continuous star formation model with
$E(B-V)=0$ ( 0 Gyr point is out of the range of the figure) and 1. We classified 
the galaxies at the region filled with yellow as the 'quiescent galaxies' 
($i'-K>3.0\quad\&\quad K-[4.5\mu$m] $<0.5$), at the
region filled with light cyan as the 'dusty starburst galaxies' ($K-[4.5\mu$m]
$>0.5$) and non-colored region as the 'normal star-forming galaxies' ($i'-K<3$
and $K-[4.5\mu$m] $<0.5$). In the GOODS-N field, $F775W$-band of {\it HST} ACS with
the detection limit of $i_{\rm 775}=25.6$ (5$\sigma$ limit) is used. Therefore
we plot the galaxies fainter than $K_{\rm AUTO}=23.6$ with open symbols in
GOODS-N. } 

\label{fig:ik45-n} 
\end{center} 
\end{figure*}

\begin{figure*}[ht] 
\begin{center} 
\leavevmode
\epsfxsize=16cm\epsfbox{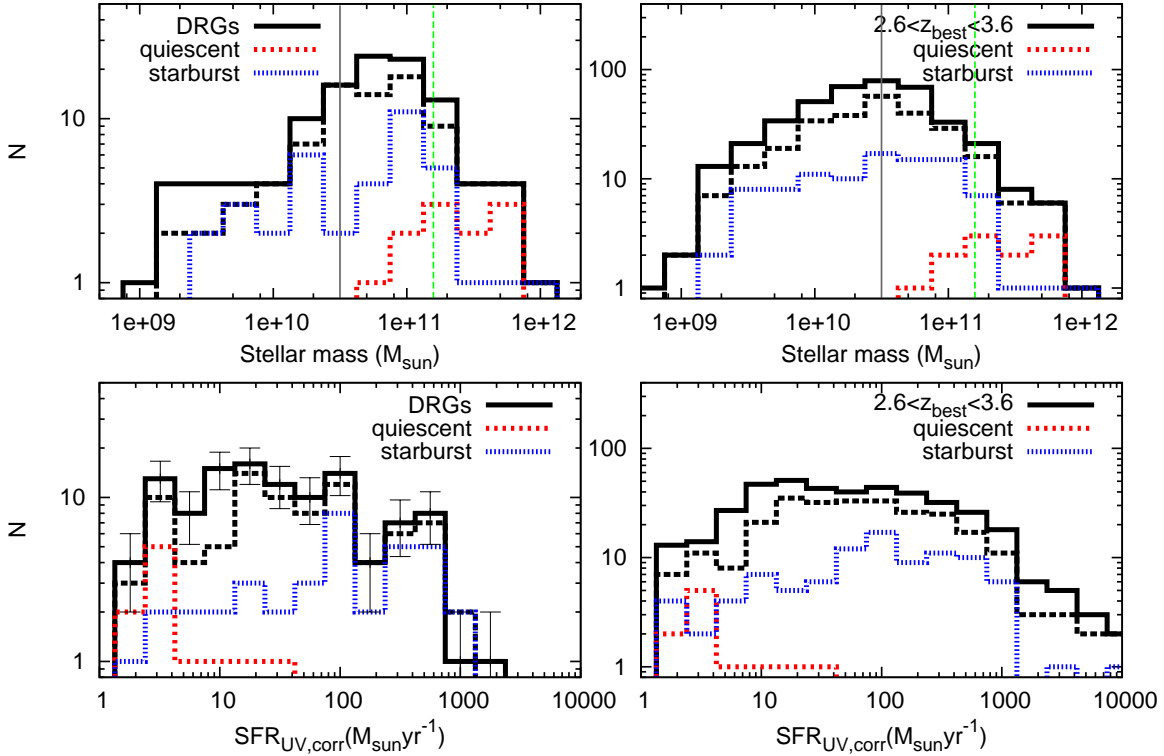}

\caption[sfr sm dist]{ {\it Top left} panel; The stellar mass distribution of
the DRGs at $2.6<z_{\rm best}<3.6$. The solid black line is the stellar mass
distribution of all the DRGs at $2.6<z_{\rm best}<3.6$ and the dashed black
line is that of the DRGs also detected in [4.5$\mu$m], which are plotted in
Fig.~\ref{fig:ik45-n}. The red and blue dashed lines are the stellar mass
distribution of the DRGs classified as the quiescent galaxies and the dusty
starburst galaxies, respectively. The gray solid line is the stellar mass limit
for typical $K$-selected galaxies with $K=24$ at $z=3.1$. The green dashed line
is that for the quiescent galaxies, which corresponds to galaxies with burst
like star formation history with the age of 0.5-2 Gyr old, with $K=23$ at
$z=3.1$. {\it Bottom left} panel; The SFR$_{\rm UV,corr}$ distribution of the
DRGs. {\it Top right} and {\it Bottom right} panels; The Stellar mass and
SFR$_{\rm UV,corr}$ distribution of the K-selected galaxies at $2.6<z_{\rm
best}<3.6$.} 
\label{fig:sfr-sm-hist} 
\end{center} 
\end{figure*}

\subsection{Star Formation Rates} 

\indent We estimated SFRs for the galaxies
using extinction corrected rest-frame UV-light.  We converted 1500--2800 \AA\
luminosities to SFR$_{\rm UV}$ values following \citep{1998ARA&A..36..189K}: 
\begin{equation*} 
{\rm SFR}_{\rm UV}(M_{\odot}~{\rm yr}^{-1})=1.4\times 10^{-28} L_{\nu}({\rm erg~s^{-1}
Hz^{-1}}).  
\end{equation*}
The conversion factor $1.4\times 10^{-28}$ was obtained assuming a Salpeter IMF
over the stellar mass range of 0.1--100 $M_{\odot}$, a constant continuous star
formation history, and stellar age older than $10^8$ yr.  Since the typical age
of LBGs at $z\simeq3$ is estimated to be $10^8-10^9$ yr assuming constant
continuous star formation model (e.g., \citealt{2001ApJ...554..981P}), the use
of this equation in our study is appropriate. We estimate the rest-frame UV
luminosities at $\sim1900$~\AA\ from their total magnitudes in $i'$-band.
The extinction corrected SFR, SFR$_{\rm UV, corr}$ are obtained by adopting the
{\it E(B-V)} value of the best fit SED models. Note that the true conversion
factor can be smaller for younger galaxies which are not reached to the
equilibrium of the birth and death of massive stars.

\indent The SFR can also be estimated using the 24$\mu$m flux, however,
\citet{2010A&A...518L..29E} point out that using the 24$\mu$m data alone
results in significant overestimation of the SFR for galaxies with $L_{\rm
IR}>10^{12}$ $L_{\odot}$ at $z=1.5-2.5$. The limiting flux at 24$\mu$m for our
observations and high-redshift sample correspond to such a high luminosity
limit.  \citet{2010A&A...518L..29E} also reported that $L_{\rm IR}$ can be
overestimated by an order of magnitude if a luminous AGN is present.  In fact,
10 out of the 31 24$\mu$m detected galaxies at $2.6<z_{\rm best}<3.6$ are also
detected by {\it Chandra}, which indicates that contributions from AGN cannot
be ignored.  We therefore choose to use only SFR$_{\rm UV, corr}$ when
estimating SFRs for our sample. We found that 21 $K$-selected galaxies had very high SFRs (SFR$_{\rm
UV,corr}\sim1000-10000$ $M_{\odot}$ yr$^{-1}$ and $E(B-V)>0.5$), but were not
detected at 24$\mu$m.  In the following analyses, we reject these sources from
our sample, since they are likely to be low-redshift contaminants.

\indent The range, average, and median of the SFR$_{\rm UV, corr}$ of the
$K$-band detected galaxies at $z_{\rm spec}\approx3.09$ in the SSA22 field are
$\sim10-730$ $M_{\odot}$ yr$^{-1}$, $122$ $M_{\odot}$ yr$^{-1}$, and $41$
$M_{\odot}$ yr$^{-1}$, respectively. This is consistent with the sample from
\citet{2010MNRAS.401.1521M}, who studied LBGs with a similar range of stellar
mass at $z_{\rm spec}\sim3$. 

\subsection{Stellar Masses}

\indent  We estimated stellar masses for our galaxies using their best-fit
SEDs and the $K$-band total magnitude.  Our sample is nearly complete for the
galaxies with stellar masses $> 10^{10.5}$ $M_{\odot}$ independent on their SEDs. Fig.
\ref{fig:sfr-sm-hist} shows the stellar mass and SFR$_{\rm UV,corr}$
distributions of the $K$-selected galaxies and DRGs at $2.6<z_{\rm best}<3.6$
in the SSA22 field.

\indent Since we study the galaxies at $z\approx3.09$, some systems will have
contributions from the [O$_{\rm III}$]$\lambda\lambda$5007 emission line
falling in the $K$-band. Assuming the conversion factor of H$\alpha$ line
luminosity to SFR from \citet{1998ARA&A..36..189K}, the line ratio of H$\alpha$
to H$\beta$ as 2.75 (\citealt{1989agna.book.....O}), H$\beta$ to [O$_{\rm
III}$] as $\sim0.3-3$ (\citealt{2006ApJS..164...81M}), and moderate extinction
$E(B-V)\Simleq0.1$, the expected flux of [O$_{\rm III}$] emission line of a
galaxy with SFR $\sim 50$ $M_{\odot}$ yr$^{-1}$ at $z\sim3$ is
$\sim10^{-16}-10^{-17}$ erg cm$^{-2}$ s$^{-1}$.  Such a [O$_{\rm III}$]
intensities may provide a small non-negligible contribution ($\Simleq0.5$ mag)
to the measured $K$-band fluxes for $z=3.0-3.2$ galaxies near the faint end of
our sample ($K\simeq23-24$). Empirically, the rest-frame equivalent width of
[O$_{\rm III}$]$\lambda\lambda$5007 emission line have been observed with
values up to $\sim$ 1000 \AA\quad (\citealt{2007ApJ...668..853K};
\citealt{2011ApJ...743..121A}).  However, the actual strength of [O$_{\rm
III}$] emission lines for the various type of galaxies at $z\sim3$ remains
uncertain and it is beyond the scope of this paper to robustly evaluate the
effect for each individual object. 

\subsection{Rest-Frame UV-to-NIR Colors}

\indent Figure \ref{fig:ik45-n} displays the $i'-K$ versus $K-[4.5\mu$m]
color-color diagram for $K$-selected galaxies with $2.6 < z_{\rm best} < 3.6$
in the SSA22 and GOODS-N fields.   This diagram has been used in past studies
to distinguish between star-forming and quiescent galaxies at $z >2$.  For
example, \citet{2005ApJ...624L..81L} and \citet{2006ApJ...640...92P} studied
$z>2$ DRGs in the Hubble Deep Field-South (HDF-S) and GOODS-South field,
respectively.  For the SSA22 sample, we show the 268 $K$-selected galaxies at
$2.6 < z_{\rm best} < 3.6$ that are detected in the 4.5$\mu$m and do not suffer
from source blending with nearby sources. The detection limit at 4.5$\mu$m
after applying aperture corrections is $\approx$ 23.9 (5$\sigma$ limit). \\
\indent We classified galaxies by comparing their $i'-K$ versus $K-[4.5\mu$m]
colors with simple synthetic models. We classified galaxies with
$i'-K>3.0\quad\&\quad K-[4.5\mu{\rm m}]<0.5$ as quiescent galaxies, since these
colors are consistent with those expected for galaxies with single-burst ages
$>0.5$~Gyr.  Galaxies with $K-[4.5\mu$m]$>0.5$ were classified as heavily dust
obscured starburst galaxies, since these colors are consistent with those
expected for galaxies with constant continuous star formation history with
$E(B-V)\sim 1$.  Finally, the star-forming galaxies with moderate dust
obscuration are expected to distribute throughout the region with $i'-K<3.0$ and
$K-[4.5\mu$m] $<0.5$.  We indicate the DRGs and LBGs (
\citealt{2000ApJ...532..170S}) in Fig.~\ref{fig:ik45-n}.  The DRGs span a wide range
of colors and dominate the population of the reddest galaxies in $i'-K$.  On the other hand, 
LBGs are typically blue in $i'-K$ and correspond to star-forming galaxies
with moderate extinction.  There are some very red objects in $K-[$4.5$\mu$m]
$\Simreq1.5$ which could be AGNs with power-law spectra in the NIR (e.g.,
\citealt{2007ApJ...660..167D}; \citealt{2009ApJ...699.1354Y}).  

\indent We carefully checked the observed SEDs of the quiescent galaxy
candidates, since their red colors could be due to the presence of strong
[O$_{\rm III}$]$\lambda\lambda$5007 emission lines (see discussion above).
Indeed 4 out of the 15 candidates show clear excesses in the $K$-band only;
they are typically $\sim0.3$ mag redder in $H-K$ and $\sim0.4$ mag bluer in
$K-[3.6$$\mu$m] than the average colors of the remaining 11 objects. All of
these 4 galaxies are fainter than $K=23$. The SEDs of the remaining 11 objects,
on the other hand, are well fitted by the models of quiescent galaxies with
ages of 0.5--2~Gyr at $z=3.1$.  If we consider only galaxies with $K<23$, we
find a corresponding $2.6<z_{\rm best}<3.6$ quiescent galaxy surface density of
$0.10\pm0.03$ arcmin$^{-2}$ in SSA22.  All of these quiescent galaxies satisfy the DRG color selection criterion in
$J-K$, suggesting the presence of significant Balmer or 4000~\AA\quad
breaks.  On the other hand, $\approx20\%$ of the DRGs with $K<23$ at $2.6<z_{\rm
best}<3.6$ in SSA22 lie in the color range of quiescent
galaxies. 

\indent There is one quiescent galaxy that is detected in 24$\mu$m. However,
the SED of this object is also well fitted by a model of passive evolution with
an age of 1.7 Gyr at $z\approx3.09$, with a slight excess over the model at
$>5.8$$\mu$m.  The observed light at wavelengths shorter than 4.5$\mu$m is
dominated by the old stellar component, while the mid-infrared emission maybe
due to dust heated by an AGN. For the remaining 10 objects without 24$\mu$m
detections, we stacked their 24$\mu$m images and found no significant stacked
detection. The corresponding upper limit of the 24$\mu$m flux for the stacked
sources is 17 $\mu$Jy $(2\sigma)$.

\indent Figure \ref{fig:sfr-sm-hist} shows the stellar mass and SFR$_{\rm
UV,corr}$ distributions of the quiescent galaxies.  Note that the SFR$_{\rm UV,
corr}$ of quiescent galaxies are just upper limits, since these are obtained
with the assumption that their UV light is dominated by massive young stars
(see discussion in Section 4.1).  The SFR$_{\rm UV, corr}$ upper limits of the
quiescent galaxies are typically less than 10 $M_{\odot}$~yr$^{-1}$, which is
significantly lower than the average value for LBGs at $z\sim3$ (see, e.g.,
\citealt{2003ApJ...588...65S} and \citealt{2010MNRAS.401.1521M} for other
samples of LBGs).

\indent The stellar masses of the quiescent galaxies span the
range of $10^{10.8-11.7}$ $M_{\odot}$, comparable to those of local
massive early-type galaxies. On the other hand, $\approx 20$\% (9/43) of the
$K$-selected galaxies with stellar mass $>10^{11}$ $M_{\odot}$ at $2.6<z_{\rm
best}<3.6$ in SSA22 are classified as the quiescent sources.  

\indent The deficit of quiescent galaxies below $\sim10^{11}$ $M_{\odot}$
is likely to be a selection effect.  Our burst models for galaxies with ages in the range of 0.5--2~Gyr
have larger stellar mass to light ratios than typical $K$-selected galaxies at
$z\sim3$.  Therefore, the limit of $K$=23 corresponds to $\sim10^{11}$
$M_{\odot}$ (The dashed green line in Fig.~\ref{fig:sfr-sm-hist}).  As
discussed above, while we attempted to select the quiescent galaxies as faint as
$K=24$, all of those with $K=23-24$ are likely to be [O$_{\rm III}$] emitters.

\indent The presence of these quiescent galaxies in the SSA22 field is
conspicuous. With the same color criterion and detection limits, we found {\it
no} such quiescent galaxies at $2.6<z_{\rm best}<3.6$ in the MODS GOODS-N
sample.  Furthermore, few DRGs in GOODS-S field satisfy our selection criterion for quiescent
galaxies (\citealt{2006ApJ...640...92P}) despite having a broader redshift range 
for selection.  If we correct the number of the
foreground/background galaxies adopting the 1.8 times density excess of the
protocluster field (Table \ref{table:table-etc}), the fraction of the quiescent
galaxies among the $K$-band selected galaxies with stellar mass $>10^{11}$
$M_{\odot}$ is expected to be at most $\approx 50\%$ (i.e., 9/19).

\indent On the other hand, 96 of the $K$-selected galaxies at $2.6<z_{\rm
best}<3.6$ were classified as dusty starburst galaxies
($K-[4.5\mu$m]$>0.5$). At stellar mass $>10^{11}$ $M_{\odot}$, they occupy
$\approx 30$\% of the sample (14/43).  With the field correction, this implies a fraction of
$\approx20\%$ and a surface density of the dusty starburst galaxies
of 0.86$\pm0.09$ arcmin$^2$, which is 1.4 times of that in the GOODS-N field.
18 of the whole dusty starburst galaxies are detected at 24$\mu$m, 
supporting the idea that they actually suffer from heavy dust obscuration.  The
median SFR$_{\rm UV,corr}$ and stellar mass of the dusty starburst galaxies
are $92$ $M_{\odot}$ yr$^{-1}$ and $3.6\times10^{10}$ $M_{\odot}$,
respectively. 

\indent We also stacked the 24$\mu$m images of the other 36 relatively isolated
dusty starburst galaxies that were not individually detected.  We find a $>3
\sigma$ detection in the stacked image with corresponding mean flux of 13
$\mu$Jy.  When limiting our stacking to a sample of 18 dusty galaxies that also
satisfy the DRG color criterion, we also obtained significant $7.5\sigma$
detection with a mean flux of 48 $\mu$Jy.

\begin{figure} 
\begin{center}
\epsfxsize=9.5cm\epsfbox{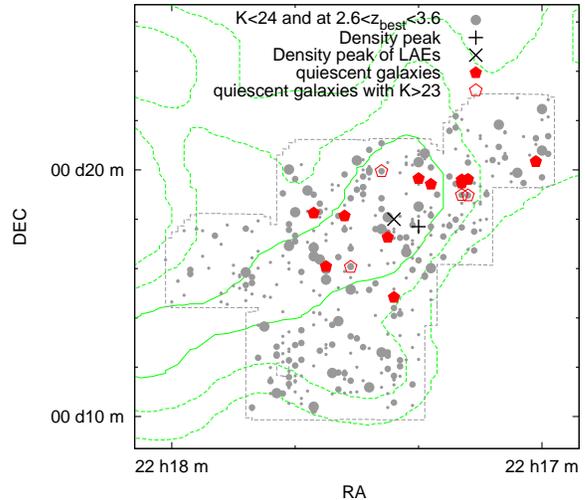}
\caption{The sky distribution of the quiescent galaxies. The red filled
pentagons are the quiescent galaxies selected as $i'-K>3.0 \quad\& \quad
K-[4.5\mu$m]$<0.5$ and $K<23$ at $2.6<z_{\rm best}<3.6$. The open red pentagons
are the $K$-selected galaxies at $2.6<z_{\rm best}<3.6$ which satisfy the color
criterion of the quiescent galaxies but with $K>23$. The gray filled circles
are the galaxies with $K<24$ at $2.6<z_{\rm best}<3.6$. The size of the symbols
indicates the stellar masses of the galaxies, classified as $M_{\rm
star}<10^{10.5}$ $M_{\odot}$ (small), $10^{10.5}$ $M_{\odot} <{\rm M}_{\rm
star} <10^{11}$ $M_{\odot}$ (medium) and $M_{\rm star}>10^{11}$ $M_{\odot}$
(large). The green contours are 1$\sigma$, 1.5$\sigma$, 2$\sigma$ density
levels of LAEs (\citealt{2004AJ....128.2073H}). The cross shows the density
peak of LAEs from \citet{2012AJ....143...79Y} and the x-mark shows the density
peak of the $K$-selected galaxies at $2.6<z_{\rm best}<3.6$ estimated in Section 3.3. } 
\label{fig:dis-passive} 
\end{center} 
\end{figure}

\begin{figure} 
\begin{center}
\epsfxsize=8cm\epsfbox{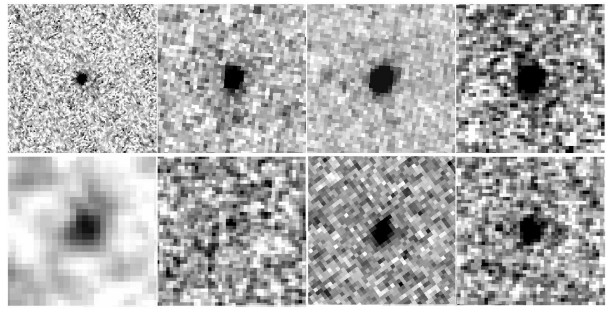}
\caption{The NIR images of the two quiescent galaxies covered by {\it HST} WFC3
IR channel. The size of each image are 5.0$''$ square. The top panels are the
{\it HST} ACS $F814W$-, WFC3 $F110W$-, $F160W$-, and MOIRCS $K$-bands images of
CXOSSA22 J2217254-001717 from left to right. The bottom panels are the images
of another quiescent galaxy which only covered by $F160W$-band of {\it HST}
WFC3. The panels are Suprime-Cam $i'$-, MOIRCS $J$-, {\it HST} WFC3 $F160W$-
and MOIRCS $K$-bands from left to right.} \label{fig:hst-passive} 
\end{center}
\end{figure}

\begin{figure} 
\begin{center}
\epsfxsize=8.5cm\epsfbox{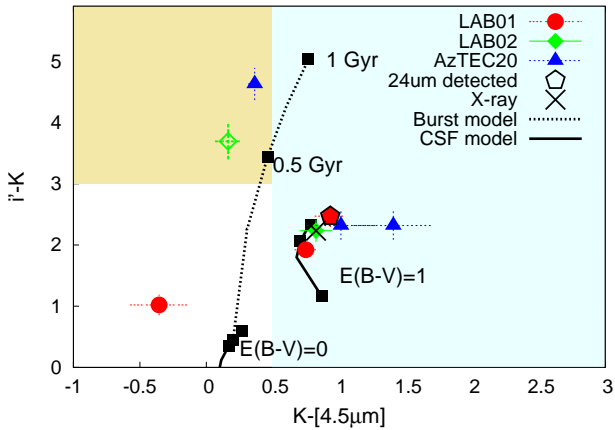}
\caption{ $i'-K$ v.s. $K-[4.5\mu$m] two color diagram similar to Fig.
\ref{fig:ik45-n} but for the counterparts of two LABs and an ASTE/AzTEC sub-mm
source. The red filled pentagons are the $K$ and [4.5$\mu$m] detected
counterparts of LAB01, the green filled diamond is that of LAB02 and the blue
filled triangles are those of AzTEC 20. The green open diamond is the
counterpart of LAB02 with slightly small photometric redshift. The black
pentagon is the galaxy detected in 24$\mu$m and the black cross is the X-ray
source. The models plotted here are same as the Fig.~\ref{fig:ik45-n}. }
\label{fig:ik45-LAB} 
\end{center} 
\end{figure}

\section{Discussion}

\subsection{Quiescent Galaxies}

\indent The presence of quiescent galaxies at high-redshift is very important for
studying the history of massive galaxy formation in clusters.  In the SSA22
protocluster, we detected 11 quiescent
galaxies based on their $i'-K$ and $K-[4.5\mu$m] colors (Fig.
\ref{fig:ik45-n}). All of these galaxies are also DRGs, showing strong Balmer or 4000~\AA\ breaks
between the observed $J$- and $K$-bands, large stellar masses, and low SFRs.  Fig.
\ref{fig:dis-passive} shows the sky distribution of these quiescent galaxies.  They
are concentrated near the density peak of protocluster LAEs and
$K$-selected galaxies at $2.6<z_{\rm best}<3.6$.  Such significant
clustering of quiescent galaxies at $z\Simreq3$ has not been clearly shown before.

\indent One of the quiescent galaxies in our sample was identified as a $z=3.1$
LAE detected in the {\it Chandra} X-ray catalog, CXOSSA22 J221725.4+001717
(\citealt{2004AJ....128.2073H}; \citealt{2009ApJ...691..687L}) at $z_{\rm
spec}=3.12$. The quoted redshift indicates that it is located at the outskirt
of the protocluster or may belong to the field.  The observed SED for this
galaxy is well fitted by an exponentially decay SFR model with $\tau=0.1$, age
of 0.72~Gyr, and $E(B-V)=0.1$.  We measured a stellar mass of
$1.6\pm0.2\times10^{11}$ $M_{\odot}$, the SFR$_{\rm UV}<9.8\pm0.7$ $M_{\odot}$
yr$^{-1}$, dust-corrected SFR$_{\rm UV, corr} < 18.3\pm9.2$ $M_{\odot}$
yr$^{-1}$, and specific SFR is $\Simleq10^{-10}$ Gyr$^{-1}$.  The SED therefore
indicates that the galaxy is quiescent.  There are no significant AGN features
based on rest-frame UV to optical indicators, as is often the case with massive
galaxies at high redshift (\citealt{2009ApJ...699.1354Y}).  However, an AGN
component may be responsible for its Ly$\alpha$ emission.\\
\indent It is also interesting to study the structural properties of the quiescent galaxies. Although a
detailed morphological study of $z \sim 3$ galaxies is beyond the MOIRCS
capabilities,  we found that two of the quiescent galaxies were observed in
existing {\it HST} data in archive.  Fig.~\ref{fig:hst-passive} shows the {\it
HST} ACS, WFC3 and the MOIRCS images of these two quiescent galaxies. Top
panels of the figure are the images of CXOSSA22 J221725.4+001717 discussed
above.  The object shown in the bottom panels is not spectroscopically
confirmed but does show an excess at the 3$\sigma$ significance in the $NB497$
narrow-band Ly$\alpha$ emission. CXOSSA22 J221725.4+001717 is well resolved in
the $F160W$-band.  We fit the image with a S\'ersic model
(\citealt{1968adga.book.....S}) using GALFIT software
(\citealt{2002AJ....124..266P}).  We obtained S\'ersic profile parameters of
$r_e=0.91\pm0.03$ kpc and $n=4.51\pm0.51$.  Such a profile is ``de~Vaucouleurs
like'' and significantly compact, resembling the red compact massive galaxies
observed at $z>2$ in other recent papers (e.g., \citealt{2008ApJ...677L...5V};
\citealt{2009ApJ...700..221K}; \citealt{2012ApJ...759L..44G}). 

\indent \citet{2012ApJ...750...93P} and \citet{2012ApJ...744..181Z} reported
the enhanced structural evolution of early-type galaxies in protoclusters at
$z=1.62$ and $z=2.16$, respectively.  Due to small number statistica,
confirming similar enhancements for the SSA22 protocluster would require
additional high-resolution images over those currently available.

\indent It is interesting to note that some of the quiescent galaxies are
likely to be members of ``multiple merging'' systems (U12).  U12 reported
that about 40\% of the LABs in the SSA22 field have multiple components
in their extended Ly$\alpha$ emission.  These components are therefore likely to be physically associated and
will merge with each other to form massive galaxies.   The most significant
cases, LAB01 and LAB02 (\citealt{2004AJ....128..569M}) respectively contain 5 and 6 massive
components at $z_{\rm phot} \sim 3.1$ in their
Ly$\alpha$ halos, which extended to scales larger than $\sim 150$ kpc.  Furthermore, a sub-mm
detected source AzTEC20 (\citealt{2009Natur.459...61T}) is also composed of 6 components
with very red colors at $z_{\rm phot} \sim 3.1$ in the beam size of ASTE/AzTEC,
$d \sim 20''$.  The summed total stellar mass of the counterparts are larger than a
few times 10$^{11}$ $M_{\odot}$ for each object. 

\indent Figure \ref{fig:ik45-LAB} shows the $i'-K$ and $K-[4.5\mu$m] colors for
the $K$- and [4.5$\mu$m]-band detected counterparts of LAB01, LAB02 and
AzTEC20.  It appears that there are quiescent galaxies in LAB02 and AzTEC20,
although the object in LAB02 has a photometric redshift slightly less than 2.6.
The stellar masses of these quiescent galaxies are $7.3\pm0.5\times10^{10}$
$M_{\odot}$ for the object in LAB02 and $5.6\pm0.2\times10^{11}$ $M_{\odot}$
for the object in AzTEC20.  The latter is already comparable to the most
massive early-type galaxies in the present-day universe.  There are also the
galaxies with the colors consistent with the dusty starburst galaxies with
$>5\times 10^{10}$ $M_{\odot}$ in the same groups.  The quiescent galaxies in
LAB02 and AzTEC20 may therefore experience major and minor mergers as the halo
grows.  At least some of the quiescent galaxies in the SSA22 protocluster have
not been completely assembled and will experience further evolution.  

\subsection{The Dust Obscured Star Forming Galaxies}

\begin{figure} 
\begin{center}
\epsfxsize=9.5cm\epsfbox{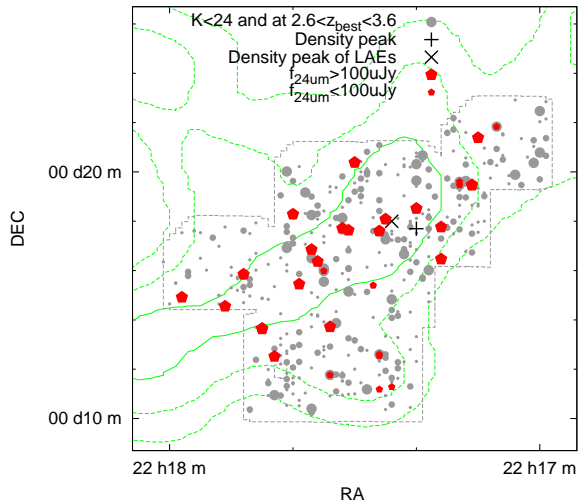}
\caption{Similar to Fig.~\ref{fig:dis-passive} but for the $K$-selected
galaxies detected in 24$\mu$m at $2.6<z_{\rm best}<3.6$. The large red filled
pentagons are the galaxies detected with $f_{24\mu{\rm m}}>100$ $\mu$Jy,
nominal detection limit in 24$\mu$m. The small filled pentagons are the
galaxies with 40 $\mu$Jy$<f_{24\mu{\rm m}}<$100 $\mu$Jy. } \label{fig:dis-24}
\end{center} 
\end{figure}

The enhancement of the dusty starburst activity in the SSA22 protocluster was previously
reported based on observations at sub-mm wavelengths
(\citealt{2007ApJ...655L...9G}; \citealt{2009Natur.459...61T};
\citeyear{2010ApJ...724.1270T}; \citeyear{2013MNRAS.430.2768T}).  Similarly, the
density excess of red galaxies like DRGs and/or HEROs  was previously reported by U12.  One of the goals of this paper,
has been to investigated the rest-frame UV to NIR colors of the protocluster galaxies to
measure the density enhancement of the dusty starburst galaxies (Section 4.3).  A large
fraction of the DRGs and/or HEROs at $2.6<z_{\rm best}<3.6$ indeed show colors
consistent with such dusty objects.  We also found a density enhancement of
24$\mu$m detected galaxies at $2.6<z_{\rm best}<3.6$, whose sky distribution
is displayed in Fig.~\ref{fig:dis-24}.

\indent It remains uncertain whether 24$\mu$m fluxes of the $K$-selected
galaxies at $2.6<z_{\rm best}<3.6$ are typically from dusty starburst or AGN
activity.  Many sources detected at 24$\mu$m have the colors consistent with
dusty starburst galaxies (Fig.~\ref{fig:ik45-n}) and two thirds of these
sources have SFR$_{\rm UV, corr}$ larger than 50 $M_{\odot}$ yr$^{-1}$. The
entire sample of 24$\mu$m detected galaxies span a wide range of SFR from 3 to
10,000 $M_{\odot}$~yr$^{-1}$, with a median value of 82 $M_{\odot}$ yr$^{-1}$.
For comparison, their median stellar mass is $8.2\times10^{10}$ $M_{\odot}$. 

\indent Three of the $K$-selected galaxies detected at 24$\mu$m that have $2.6<z_{\rm
best}<3.6$ are located within the beam size ($d\sim$ 20$''$) of the
ASTE/AzTEC sub-mm sources (\citealt{2009Natur.459...61T}).  If we assume the IR
SEDs are similar to local starburst galaxies (\citealt{2002ApJ...576..159D}), the
implied 1.1~mm flux of the galaxies with $f_{24\mu{\rm m}}=100$ $\mu$Jy would be 1--4
mJy, which is comparable to the 3$\sigma$ detection limit of the ASTE/AzTEC
observations (\citealt{2009Natur.459...61T}). 

\begin{figure} 
\begin{center}
\epsfxsize=8cm\epsfbox{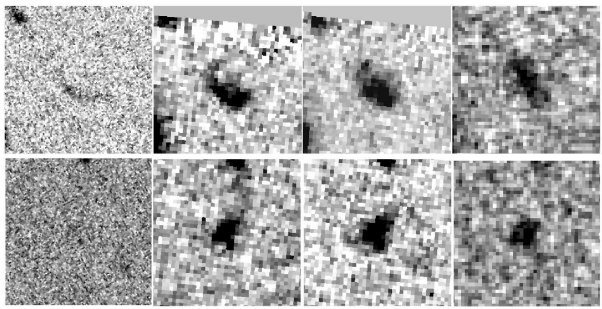}
\caption{Similar to Fig.~\ref{fig:hst-passive} but for the 24$\mu$m detected
galaxies at $2.6<z_{\rm phot}<3.6$. The panels are the {\it HST} ACS $F814W$-,
WFC3 $F110W$-, $F160W$-, and MOIRCS $K$-bands images from left to right. }
\label{fig:hst-24det} 
\end{center} 
\end{figure}

\indent Two of the $K$-selected galaxies detected in 24$\mu$m that have $2.6<z_{\rm
best}<3.6$ happen to be covered by existing WFC3 $F110W$- and
$F160W$-bands images. Fig.~\ref{fig:hst-24det} shows their ACS, WFC3 and MOIRCS
$K$-bands images.  They clearly show extended and clumpy structures, which are
quite different from quiescent red galaxies (see above).  The structural properties of
the six LBGs in the SSA22 protocluster were investigated by
\citet{2009MNRAS.398.1915M} and \citet{2011A&A...528A..88G}.  Compared with
these LBGs, which also have SFR of $>50$ $M_{\odot}$ yr$^{-1}$, stellar mass of
$10^{10.3-11.3}$ $M_{\odot}$, and relatively compact size ($r_e=1-2.5$ kpc),
the two 24$\mu$m-detected objects have more extended and diffuse structures.

\subsection{Active Galactic Nuclei}

\begin{figure} 
\begin{center}
\epsfxsize=9.5cm\epsfbox{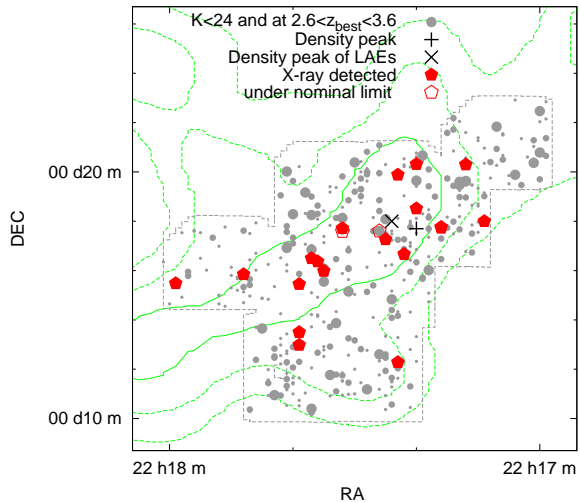}
\caption{Similar to Fig.~\ref{fig:dis-passive} but for the $K$-selected
galaxies detected in {\it Chandra} X-ray at $2.6<z_{\rm best}<3.6$. The red
filled pentagons are the galaxies detected in {\it Chandra} full- and/or soft-
and/or hard-bands above the nominal detection limits. The open pentagons are
the X-ray detected galaxies but their X-ray fluxes are fainter than the nominal
detection limits in the SSA22 field. Note that the M5 field was not covered with
{\it Chandra} data. } \label{fig:dis-x} 
\end{center} 
\end{figure}

\indent Figure \ref{fig:dis-x} shows the sky distribution of the $K$-selected
galaxies at $2.6<z_{\rm best}<3.6$ that are detected in the {\it Chandra}
0.5--8 keV band.  Given the inferred luminosity limits over this redshift
range, $L_{\rm 0.5-8~keV} > 3 \times 10^{42}$~ergs~s$^{-1}$, these X-ray
detected sources are almost certainly AGN.  The objects detected at full (0.5-8
keV), soft (0.5-2 keV), and hard (2-8 keV) bands are plotted in
Fig.~\ref{fig:dis-x}.  The X-ray sources are clustered around the density
peaks of the LAEs and $K$-selected galaxies at $2.6<z_{\rm best}<3.6$.  Note
that the {\it Chandra} data is relatively shallow over the M5 field.   We
therefore study the X-ray properties of galaxies within the 99.8 arcmin$^2$
area remaining after the exclusion of the M5 field.  In this section, we
discuss the number density of the X-ray sources, which are detected above the
nominal detection limit in the SSA22 field (listed in Section 2). As seen in Fig.~\ref{fig:ik45-n}, only a few of the X-ray detected
sources have red $i'-K$ and blue $K-[4.5\mu$m] colors.  This indicates that
their $K$-band emission is not significantly affected by the [O$_{\rm
III}$]$\lambda\lambda5007$ emission line shifted in $K$-band. 

\indent The surface number densities of the $K$-selected galaxies at
$2.6<z_{\rm best}<3.6$ associated with the X-ray sources are $0.19\pm0.04$
arcmin$^{-2}$ in the full-band sample, $0.14 \pm0.04$ arcmin$^{-2}$ in the
soft-band sample and $0.13\pm0.04$ arcmin$^{-2}$ in the hard-band sample.
These densities are 2.5$\pm1.0$ times (full-band), 2.4$\pm1.2$ times
(soft-band), and 3.4$\pm1.9$ times (hard-band) the corresponding X-ray source
densities in the GOODS-N field.  Since the number density of the whole sample
of $K$-selected galaxies at $2.6<z_{\rm best}<3.6$ in the SSA22 field is 1.7
times that in the GOODS-N field (Section 3.3), the X-ray AGN hosting rate (i.e.,
the fraction of galaxies hosting an AGN) in the hard-band is roughly twice that
of the field.  We note, however, that the statistical uncertainty here is
large. 

\begin{figure*} 
\begin{center} \leavevmode
\epsfxsize=13cm\epsfbox{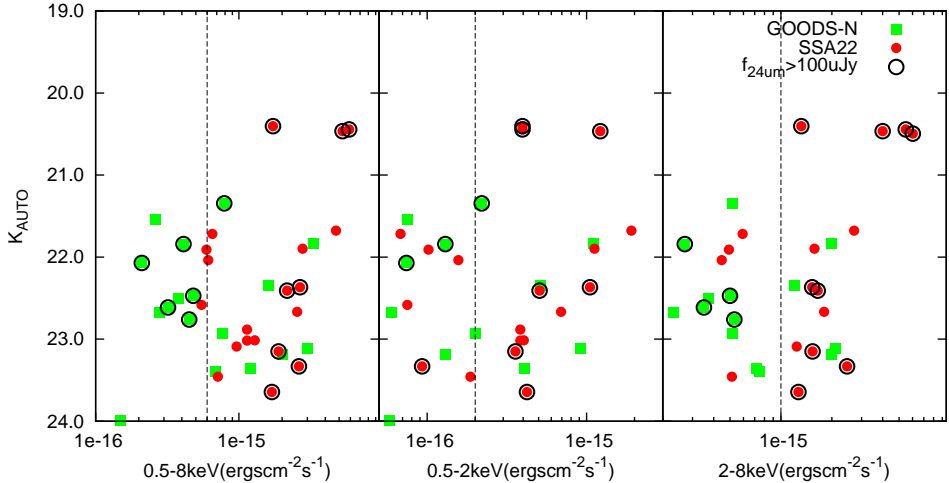}
\caption[chandra-f-s-h]{ Total magnitudes in $K$-band versus fluxes in X-ray
distributions for the $K$-selected galaxies at $2.6<z_{\rm best}<3.6$. The
figures are full- (0.5-8 keV), soft- (0.5-2 keV) and hard-band (2-8 keV) from
left to right. The red filled circles are the $K$-selected galaxies at
$2.6<z_{\rm best}<3.6$ in 99.8 arcmin$^2$ of the SSA22 field. The green filled
squares are the $K$-selected galaxies at $2.6<z_{\rm best}<3.6$ in 103.3
arcmin$^{-2}$ in GOODS-N field. The black circles are the objects with
$f_{24\mu{\rm m}}>100$ $\mu$Jy. The dashed lines in each panel are the nominal
detection limits of {\it Chandra} data in the SSA22 field. }
\label{fig:chandra-f-s-h} 
\end{center} 
\end{figure*}

\indent We plot the $K$-band magnitudes versus the X-ray flux of the
$K$-selected galaxies at $2.6<z_{\rm best}<3.6$ in Fig.
\ref{fig:chandra-f-s-h}.  In the GOODS-N field, there are only a few objects
brighter than the nominal detection limits for the {\it Chandra} observations
in SSA22.  About half of the X-ray detected objects in SSA22 are also detected
at 24$\mu$m ($>100$ $\mu$Jy).  The AGNs in the SSA22 protocluster seem to be
more powerful and more dust attenuated than those in the general field at
similar redshifts. The 24$\mu$m detection rate is largest for the hard-band
sample, which is consistent with AGN having large dust attenuation in the
protocluster. The more luminous X-ray objects are observed in the SSA22 
protocluster, which also suggests the enhanced AGN activity.  

\indent There may be further contributions from a hidden population of dust
obscured AGNs that are faint in X-ray emission. \citet{2007ApJ...670..173D}
investigated 24$\mu$m sources at $z>2$ and found that the galaxies that are
more luminous than $L_{8\mu {\rm m}}>10^{11}$ $L_{\odot}$ in $\nu L_{\nu}$ show
a statistically significant X-ray detection when stacked.  In the SSA22 field,
among 31 $K$-selected galaxies at $2.6<z_{\rm best}<3.6$ detected at 24$\mu$m,
we find 10 that are also detected in the X-ray band.  Since the detection limit
in 24$\mu$m in the SSA22 field corresponds to $L_{6\mu {\rm m}}\sim10^{11}$
$L_{\odot}$ in $\nu L_{\nu}$, it is reasonable to expect that there are
additional dust obscured AGNs not detected in X-ray. Therefore, we consider the
AGN hosting rate to be a lower limit.

\indent The excess AGN hosting rate of LAEs and LBGs in the SSA22
protocluster was reported in \citeauthor{2009ApJ...691..687L} (\citeyear
{2009ApJ...691..687L}; \citeyear{2009MNRAS.400..299L}) and
\citet{2009ApJ...692.1561W}. All the galaxies at $z_{\rm spec} \approx3.09$
detected in X-ray listed in Table 1 of \citet{2009ApJ...691..687L} lie in the
MOIRCS observed area. All of them are also members of our $K$-selected sample.
A similar enhancement was also reported for LAEs and BX/MD galaxies in the HS~1700+64
protocluster at $z = 2.30$ studied by \citet{2010MNRAS.407..846D}.

\indent The excess AGN hosting rate of the galaxies in the SSA22 protocluster
may be due to the enhanced activity of AGNs and/or the presence of more massive
host galaxies in the protocluster environment.  Given that supermassive black
hole (SMBH) masses and their host galaxy stellar masses are tightly correlated
in the local Universe (e.g., \citealt{1995ARA&A..33..581K};
\citealt{1998AJ....115.2285M}), it is possible that the enhanced AGN hosting
rate is due to the presence of more massive galaxies there. 
In Table \ref{table:table-etc}, we tabulate the surface number 
density of $K$-selected galaxies over three stellar mass ranges. The density excess in
each stellar mass bin is similar to that of the whole $K$-selected galaxy sample.
Therefore, at least, the excess AGN hosting rate of the $K$-selected galaxies at
$2.6<z_{\rm best}<3.6$ is not likely to be simply due to the presence of more massive
host galaxies in the SSA22 protocluster. The hard-band X-ray detection rates are $14\pm6$\% in SSA22 and $9\pm6$\% in GOODS-N for $K$-selected galaxies with $M_{\rm star}$ $>$
$10^{11}$ $M_{\odot}$ and $2.6<z_{\rm best}<3.6$. Further observations are needed to constrain how the protocluster environment influences the AGN activities.

\subsection{$K$-selected LAEs} 

\indent A fraction of the $K$-selected galaxies at $2.6<z_{\rm best}<3.6$ show
excess in $BV-NB497$ colors.  Such excess colors are indicative of sources with
either strong Ly$\alpha$ emission at $z\approx3.09$ or [O$_{\rm II}$] emission
at $z\approx0.33$.  Note that the $BV$ magnitude represents the average
continuum flux at the wavelength of the narrow band (Hayashino et al.~2004).
In fact, 41 out of the 433 $K$-selected galaxies show excess in narrow-band
flux above $4\sigma$ of the scatter in $BV-NB497$ color at each brightness.
This criterion is less strict than that adopted to construct robust samples of
LAEs (e.g., \citealt{2004AJ....128.2073H}); however, this result provides
support that our photometric redshift measurements are reliable.  In total
there are 78 robust $z = 3.1$ LAEs from \citet{2004AJ....128.2073H} lie within
the footprint of our MOIRCS observations.  We find that 9 of these 78 LAEs are
part of our $K$-selected sample.

\indent The work by \citet{2010ApJ...724.1524O} explored the NIR properties of
LAEs in the SXDS field and is suitable to compare with our results.  In the
SXDS field, only 5 out of the 224 LAEs at $z=3.09$ in 2340 arcmin$^2$ area were
detected with $K<24$ (3$\sigma$ limit).  Note that the narrow-band filter used
in their study, $NB507$, has slightly different central wavelength but similar
bandwidth as that of $NB497$. The narrow-band detection limit in SXDS varies
from $NB507=25.1$ to $25.5$ (5$\sigma$ limit) (\citealt{2008ApJS..176..301O}).
All the LAEs detected in $K$-band in the SSA22 field have $NB497<25.1$ while
the number of the LAEs with $NB497<25.1$ is 55 and that with $NB497<25.5$ is
69.  Therefore, the $K$-band detection rate of the LAEs in the SSA22
protocluster (9/55 or conservatively, 9/69) is 5.8--6.8 times larger than that
in the SXDS field (5/224).

\indent The properties of the protocluster LAEs detected in the $K$-band are
also different from those in the SXDS field.  The median stellar mass and
stellar mass range of $K$-selected LAEs in SSA22 are $6.1\times 10^{10}$
$M_{\odot}$ and $10^{10-11}$ $M_{\odot}$, respectively.  While in the SXDS
field, these quantities are $\approx 3.7 \times 10^{9}$ $M_{\odot}$ and
$10^{9-10.5}$ $M_{\odot}$, respectively.  It has also been reported in other
studies that the rest-frame UV selected galaxies like LBGs and LAEs in the
protocluster are typically more massive than those in the general field
(\citealt{2005ApJ...626...44S}; \citealt{2009ApJ...691..687L};
\citealt{2010MNRAS.407..846D}; \citealt{2012MNRAS.425..878M}).  Furthermore, 7
of the $K$-selected LAEs in the SSA22 protocluster satisfy the DRG color
criterion, while there are no such objects in the SXDS field.  Red LAEs have
also been identified in other studies (\citealt{2007A&A...471...71N};
\citealt{2008ApJ...674...70L}), but these sources are very rare.  We caution
that the red color could also be due to the contribution from a strong [O$_{\rm
III}$]$\lambda\lambda5007$ emission line.  As discussed above, this is likely
to be the case for two objects with $K>23$, but the remaining 7 brighter red
LAEs are unlikely to be influenced by the [O$_{\rm III}$]$\lambda\lambda5007$
emission line.  We further note that 4 of the $K$-selected LAEs are also
detected in {\it Chandra} X-ray. 

\indent The redshift evolution of the stellar populations of LAEs was reported
in \citet{2009A&A...498...13N} who found that LAEs at $z=2.2$ are redder than
those at $z=3$.  \citet{2009A&A...498...13N} also reported a higher AGN
fraction, more dust obscuration, and/or older stellar populations among the
LAEs at $z=2.2$ than those at $z=3$.  Our results indicate that the properties
of LAEs at $z=3$ are also strongly dependent upon environment.

\indent The $K$-selected LAEs in the SSA22 protocluster are not low-mass
analogs of LBGs. There are only two DRGs among the 26 LBGs at $z_{\rm
spec}\approx3.1$ that do not satisfy the robust LAE selection criteria.  In
this sense, our results are somewhat different from those in
\citet{2008ApJ...674...70L} who investigated the $R-[3.6$$\mu$m] colors of 
LAEs and LBGs at $z\sim3.1$ with deep IRAC 3.6$\mu$m data. They argued that
LAEs tend to be bluer in $R-[3.6$$\mu$m] than typical LBGs and are distributed
within the extension of the faint and blue end of the LBG population. 

\section{Conclusions} 

\indent We identified $K$-selected galaxies that are candidate members of the
SSA22 protocluster at $z=3.09$ using deep and wide NIR observations and
photometric redshift fitting. 

\indent We found the surface number density excess of the $K$-selected galaxies
at $2.6<z_{\rm best}<3.6$ in the SSA22 protocluster field over that of the
general field (e.g., from GOODS-North and GOODS-South).  We also found that
DRGs ($J-K<1.4$) and 24$\mu$m detected galaxies among the $K$-selected
population at $2.6<z_{\rm best}<3.6$ show even larger surface number density
excesses.  We identified 11 quiescent galaxies in our sample by using their
rest-frame UV to NIR colors.  No such galaxies were identified throughout the
entire GOODS-N field. Such significant clustering of the quiescent galaxies at
$z \approx 3$ is reported here for the first time. Additionally, we found a
density excess of dusty starburst galaxies selected by the color criterion,
which indicates that the SSA22 protocluster galaxies are still experiencing
enhanced star formation.  There is also a density excess of $K$-selected
galaxies at $2.6<z_{\rm best}<3.6$ detected in the X-ray bandpass.  These
sources are typically more obscured and X-ray luminous than those in the
general field.  Finally, we found an excess $K$-band detection rate for LAEs in
the protocluster.  Almost all of these LAEs are also DRGs.  Such red LAEs are
very rare in general fields.

\indent From the above, we conclude that there are already a significant
fraction of evolved quiescent massive galaxies in the SSA22 protocluster.
However, star-formation and also AGN activity is still ongoing and enhanced over the
field.  It appears that we are indeed witnessing a key formation epoch of
elliptical galaxy formation in a protocluster that will collapse to form a rich
galaxy cluster by $z = 0$.

\indent We thank MOIRCS Deep survey team for providing the photometric and
spectroscopic catalog in the GOODS-N field. We also thank  Dr. Yoichi Tamura
for providing the updated coordinates of the AzTEC sources. Our studies owe a
lot deal to the archival {\it Spitzer} IRAC \& MIPS data taken in
\citet{2009ApJ...692.1561W}, {\it Chandra} data taken in
\citet{2009ApJ...691..687L} and {\it HST} ACS and WFC3 data from proposal ID;
11636, PI; Siana. This work was supported by Global COE Program "Weaving
Science Web beyond Particle-Matter Hierarchy", MEXT, Japan. \\

\bibliographystyle{apj}
\bibliography{apj-jour,ssa22-moircs130814}
\newpage
\begin{table*}[htbp]
 \caption{Data set}
\label{table:table-nc}
	\begin{center}
\begin{tabular}{lccccr}
\hline \hline
Band & Instrument & Reference  & Depth &\\ \hline
$J$ &  MOIRCS/Subaru	& \citet{2012ApJ...750..116U}	& 24.1-24.5\footnotemark[1]& \\
$H$ &  MOIRCS/Subaru	& \citet{2012ApJ...750..116U}	& 23.6-24.0\footnotemark[1]& \\
$K$ &  MOIRCS/Subaru	& \citet{2012ApJ...750..116U}	& 24.5-25.0\footnotemark[1] & \\
$u^{\star}$  & MegaCam/CFHT & -\footnotemark[2] & 26.1\footnotemark[3] &	\\
$B$ & Suprime-Cam/Subaru	& \citet{2004AJ....128..569M}	&  26.45\footnotemark[3]&\\
$V$ & Suprime-Cam/Subaru	& \citet{2004AJ....128..569M}	& 26.5\footnotemark[3] &\\
$R$ & Suprime-Cam/Subaru	& \citet{2004AJ....128..569M}	&  26.6\footnotemark[3]&\\
$NB497$ & Suprime-Cam/Subaru	& \citet{2004AJ....128..569M}	& 26.2\footnotemark[3]&\\
$i'$ & Suprime-Cam/Subaru	& \citet{2004AJ....128.2073H}& 26.2\footnotemark[3]& \\
$z'$ & Suprime-Cam/Subaru	& \citet{2004AJ....128.2073H}& 25.5\footnotemark[3]& \\
3.6$\mu$m & IRAC/{\it Spitzer}  & \citet{2009ApJ...692.1561W}	& 24.1\footnotemark[4] & \\
4.5$\mu$m & IRAC/{\it Spitzer} & \citet{2009ApJ...692.1561W}	& 23.9\footnotemark[4] & \\
5.8$\mu$m & IRAC/{\it Spitzer} & \citet{2009ApJ...692.1561W}	& 22.5\footnotemark[4] & \\
8.0$\mu$m & IRAC/{\it Spitzer} & \citet{2009ApJ...692.1561W}	& 22.1\footnotemark[4] & \\
24$\mu$m & MIPS/{\it Spitzer}	& \citet{2009ApJ...692.1561W}	& 40-100$\mu$Jy\footnotemark[5]  & \\
0.5-8keV & {\it Chandra} 	& \citet{2009ApJ...691..687L}	& $\approx 6\times10^{-16}$ erg cm$^{-2}$ s$^{-1}$\footnotemark[6]  &  \\
0.5-2keV & {\it Chandra} 	& \citet{2009ApJ...691..687L}	& $\approx2\times10^{-16}$ erg cm$^{-2}$ s$^{-1}$\footnotemark[6] &  \\
2-8keV & {\it Chandra} 	& \citet{2009ApJ...691..687L}	& $\approx1\times10^{-15}$ erg cm$^{-2}$ s$^{-1}$\footnotemark[6] &  \\
$F160W$ & WFC3/{\it HST} 	& -\footnotemark[7]	& 26.7\footnotemark[8] &  \\
\hline
\footnotetext[1]{ 5$\sigma$ detection limiting magnitude in 1.1$''$ diameter aperture. }
\footnotetext[2]{ P. I. Cowie (in CFHT archive)}
\footnotetext[3]{ 5$\sigma$ detection limiting magnitude in 2.0$''$ diameter aperture on the images smoothed so that the PSF sizes are matched to be 1.0$''$ in FWHM. }
\footnotetext[4]{ 5$\sigma$ detection limiting magnitude corrected to the flux in 1.1$''$ diameter aperture on the images with the PSF sizes of 0.5$''$. Here, we adopt the average aperture correction factor of the relatively isolated sources. }
\footnotetext[5]{ 3$\sigma$ detection limit of the total flux in the 107.8 arcmin$^2$ field also observed with MOIRCS.  }
\footnotetext[6]{ The nominal detection limit in the 99.8 arcmin$^2$ field also observed with MOIRCS }
\footnotetext[7]{ P. I. Siana, Proposal ID; 11636 (in {\it HST} archive) }
\footnotetext[8]{ 5$\sigma$ detection limiting magnitude in 0.6$''$ diameter aperture }
\end{tabular} 
     \label{tab:tab1}
\end{center}
\end{table*}
\begin{table*}[htbp]
 \caption{Number counts of the $K<24$ samples }
\label{table:table-nc}
	\begin{center}
\begin{tabular}{lccr}
\hline \hline
Selection & $N_{\rm obj}$\footnotemark[1] & $N_{\rm obj}$ with $2.6<z_{\rm best}<3.6$ ($2.6<z_{\rm spec}<3.6$)\footnotemark[2] &		\\ \hline
ALL		&	4520 	&	433(49)	&  \\
DRGs				&	364		&	116(8)	& \\
\hline
\footnotetext[1]{ Number of the objects with $K<24$ in the MOIRCS observed field in the SSA22 field. }
\footnotetext[2]{ Number of the galaxies with $K<24$ and $2.6<z_{\rm best}<3.6$ ($2.6<z_{\rm spec}<3.6$). }
\end{tabular} 
\end{center}
\end{table*}
\begin{table*}[htbp]
 \caption{Number counts of the characteristic $K$-selected galaxies at $2.6<z_{\rm best}<3.6$ }
\label{table:table-etc}
	\begin{center}
\begin{tabular}{lccccr}
\hline \hline
Classification   & Criterion & $N_{\rm obj}$\footnotemark[1] &	Number density & Ratio\footnotemark[2] &	\\ 
	  & 		& 							 &	(arcmin$^{-2}$) & (SSA22 / field)	&\\ \hline
All the $K$-selected galaxies &	$K<24$	&	433(49)	& 3.87$\pm0.19$ & 1.7$\pm$0.1 &\\
	 &	$M_{\rm star}$ $>$ $10^{11}$ $M_{\odot}$	&	43(11)	& 0.38$\pm0.06$ & 1.8$\pm$0.5 &\\
	 &  10$^{10.5}$ $M_{\odot}$ $<$ $M_{\rm star}$ $<$ $10^{11}$ $M_{\odot}$	&	116(7)	& 1.03$\pm0.10$ & 1.6$\pm$0.2 &\\
	 &	 $M_{\rm star}$ $<$ 10$^{10.5}$ $M_{\odot}$ 	&	274(31)	& 2.45$\pm0.15$ & 1.7$\pm$0.2 &\\
Quiescent galaxies &	$i'-K>3\quad \&\quad K-[4.5$ $\mu{\rm m}]\quad \&\quad K<23$	&	11(1)	& 0.10$\pm0.03$ & - \footnotemark[3]&\\
All of the 24$\mu m$ detected	&	 $f_{24\mu{\rm m}}>40$ $\mu$Jy	&	31(5)	& 0.29$\pm0.05$ & - &\\
24$\mu m$ detected	&	 $f_{24\mu{\rm m}}>100$ $\mu$Jy	&	22(3)	& 0.20$\pm0.04$ & 1.9$\pm$0.7 &\\
{\it Chandra} full-band detected	& $f_{\rm 0.5-8keV}>$6$\times10^{-16}$ erg cm$^{-2}$ s$^{-1}$	&	19(10)	& 0.19$\pm0.04$ & 2.5$\pm$1.0 &\\
{\it Chandra} soft-band detected	& $f_{\rm 0.5-2keV}>$2$\times10^{-16}$ erg cm$^{-2}$ s$^{-1}$	&	14(8)	& 0.14$\pm0.04$ & 2.4$\pm$1.2 &\\
{\it Chandra} hard-band detected & $f_{\rm 2-8keV}>$1$\times10^{-15}$ erg cm$^{-2}$ s$^{-1}$	&	13(8)	& 0.13$\pm0.04$ & 3.4$\pm$1.9 &\\
LAEs	& $BV-NB497>1.2$ \& $NB497<25.5$	&	9(5)	& 0.08$\pm0.03$ & 6.8$\pm2.3$\footnotemark[4] &\\
\hline
\footnotetext[1]{ Number of the $K$-selected galaxies at $2.6<z_{\rm best}<3.6$ ($2.6<z_{\rm spec}<3.6$). }
\footnotetext[2]{ Ratio of the surface number density in the SSA22 protocluster to that in the general field. All the samples but LAEs are compared with those in the MOIRCS Deep Survey in GOODS-N field (\citealt{2011PASJ...63S.379K}). The LAEs are compared with those in SXDS field (\citealt{2008ApJS..176..301O}; \citealt{2010ApJ...724.1524O}). }
\footnotetext[3]{ None of the $K$-selected galaxies at $2.6<z_{\rm best}<3.6$ in GOODS-N field satisfy the color criterion of the quiescent galaxies. }
\footnotetext[4]{ Ratio of $K$-band detection rate of the LAEs in SSA22 and in SXDS field. }
\end{tabular} 
\end{center}
     \label{tab:tab3}
\end{table*}
\end{document}